\pdfoutput=1
\documentclass{JINST}
\usepackage{amsopn}
\usepackage{lineno}
\usepackage{subfigure}
\usepackage{epstopdf}

\usepackage{SIunits}
\usepackage{color} 
\makeatletter
\color{black}
\let\default@color\current@color
\makeatother
\newcommand{\fprompt}{F$_{\mathrm{prompt}}$}
\newcommand{\fpromptmath}{\mathrm{F}_{\mathrm{prompt}}}
\DeclareMathOperator\erfc{erfc}

\title{Temperature dependence of alpha-induced scintillation in the 1,1,4,4-tetraphenyl-1,3-butadiene wavelength shifter}

\author{L.~M.~Veloce\thanks{Corresponding author.}~\footnote{Current address: {\it Department of Physics, University of Toronto, Toronto, Ontario, Canada, M5S 1A7}}~, M.~Ku\'zniak\thanks{Corresponding author.}, P.~C.~F.~Di~Stefano, A.~J.~Noble, M.~G.~Boulay, P.~Nadeau, T.~Pollmann, M.~Clark, M.~Piquemal, and K.~Schreiner \\ 
\llap \ Department of Physics, Engineering Physics \& Astronomy,\\Queen's University, Kingston, Ontario,\\Canada, K7L 3N6\\
E-mail: \email{lveloce@physics.utoronto.ca}, \email{kuzniak@owl.phy.queensu.ca}}

\abstract{Liquid noble based particle detectors often use the organic wavelength shifter 1,1,4,4-tetraphenyl-1,3-butadiene (TPB) which shifts UV scintillation light to the visible regime, facilitating its detection, but which also can scintillate on its own.
Dark matter searches based on this type of detector commonly rely on pulse-shape discrimination (PSD) for background mitigation. Alpha-induced scintillation therefore represents a possible background source in dark matter searches. The timing characteristics of this scintillation determine whether this background can be mitigated through PSD.
We have therefore characterized the pulse shape and light yield of alpha induced TPB scintillation at temperatures ranging from 300~K down to 4~K, with special attention given to liquid noble gas temperatures. We find that the pulse shapes and light yield depend strongly on temperature. In addition, the significant contribution of long time constants above ~$\sim$50~K provides an avenue for discrimination between alpha decay events in TPB and nuclear-recoil events in noble liquid detectors.}


\begin{document}


\section{Introduction}

1,1,4,4-Tetraphenyl-1,3-Butadiene (TPB) is an organic compound commonly used as a wavelength shifter in a variety of particle experiments which employ noble liquids as the detector medium, including DEAP-3600~\cite{deap_proc}, MiniCLEAN~\cite{miniclean}, DarkSide~\cite{darkside}, ArDM~\cite{ardm}, WArP~\cite{warp}, nEDM at SNS~\cite{nedm}, and MicroBooNE~\cite{microboone}. It converts the UV light emitted by the noble liquids to easier-to-detect visible light.

TPB also scintillates in response to excitation by alpha particles. A careful examination of the characteristics of TPB scintillation is critical for ultra-low background experiments in order to minimize the risk of introducing false dark matter (DM) signals. Of particular concern for ultra-low background detectors are false signals coming from alpha emitters, usually $^{222}$Rn radioactive decay daughters, embedded in detector walls or in the TPB itself.
Recoiling daughter nuclei depositing energy in the noble liquid target, and emitted in the decay back-to-back with alphas entering the TPB-coated detector wall, generate low energy events similar to the signature expected from Weakly Interacting Massive Particles (WIMPs), especially if the $\alpha$ is not tagged.
Another category of events are $\alpha$ particles interacting in the TPB, while the associated nuclear recoil is lost in the non-scintillating detector wall.
The presence of TPB affects this class of events due to its scintillation properties, and therefore deserves a detailed study at cryogenic temperatures.

\begin{figure}[!t]
\centering
\includegraphics[width=120mm,scale=1.5]{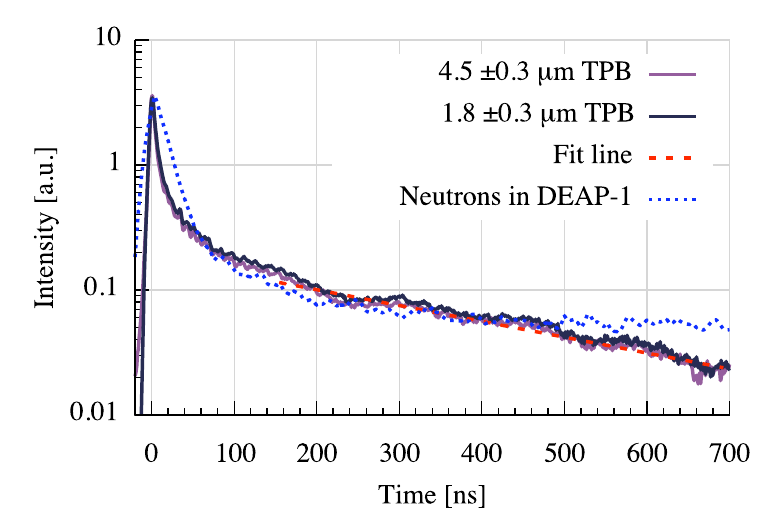}
\setlength{\belowcaptionskip}{10pt plus 3pt minus 2pt}
\caption{\small Neutron pulse shape from DEAP-1 compared with alpha-induced TPB scintillation pulse shapes at room temperature (from reference~\cite{TinaPaper}). \fprompt\ values are similar, but the long time constants are quite different ($\sim$~1.5~\micro\second\ vs. 275$\pm$10~\nano\second\ respectively). The short time constants also appear quite different, however, the error on the short time constant in this measurement is quite large. Note that TPB thickness does not affect the shape of the curve.}
\label{fig:tinagraph}
\end{figure}
 
DEAP-3600 is a single-phase liquid argon-based detector which relies on pulse shape discrimination (PSD) techniques for particle identification. In liquid argon, scintillation has two main components, of which one has a fast and one a slow decay constant, 6~ns and $\sim$~1.5~\micro\second, respectively~\cite{hitachi}. The relative amplitudes of the fast and slow components of a liquid argon scintillation pulse can be used to identify particle type. This can be quantified using a PSD parameter called the Prompt Fraction, or \fprompt.

An extensive study of alpha-induced TPB scintillation has already been completed at room temperature~\cite{TinaPaper}. Alpha-induced TPB scintillation was found to have a long time constant of 275~$\pm$~10~ns, which is very different from liquid argon time constants (see Figure~\ref{fig:tinagraph}).
Even though the scintillation in TPB was found to have an \fprompt\ value of 0.67 $\pm$ 0.03, very close to the measured \fprompt\ of $\sim$0.74 at 20~keV (electron equivalent) for nuclear recoils in DEAP-1~\cite{TinaPaper,Boulay2009}, the presence of the additional long time constant allows one to introduce a dedicated new PSD parameter based on charge fractions, with integration limits tuned to the TPB scintillation timing, in order to maximize the discrimination.

It has been recently suggested~\cite{Segreto2015} that long time constants are also present in emission from TPB excited by 128~nm VUV photons (i.e. argon scintillation light), which indicates the occurrence of ionization and triplet state generation in TPB. Still, the ratio of fast and delayed components for LAr photons and for alphas should be significantly different, as it is expected to depend on the linear energy transfer, and consequently on the particle type. This makes an accurate measurement of the TPB response to alphas even more important.

Because the scintillation properties of organic compounds can change dramatically with temperature (see for instance Ref.~\cite{Baker2008}), in this work we conducted an investigation of alpha-induced TPB scintillation in conditions relevant for cryogenic experiments, i.e. at temperatures ranging from 300~K to 4~K, with special attention given to noble liquid temperatures.

\section{Method}

\begin{figure*}[!t]
\centering
\includegraphics[trim=0cm 0cm 0cm 0cm, clip=true, totalheight=0.4\textheight, angle=0]{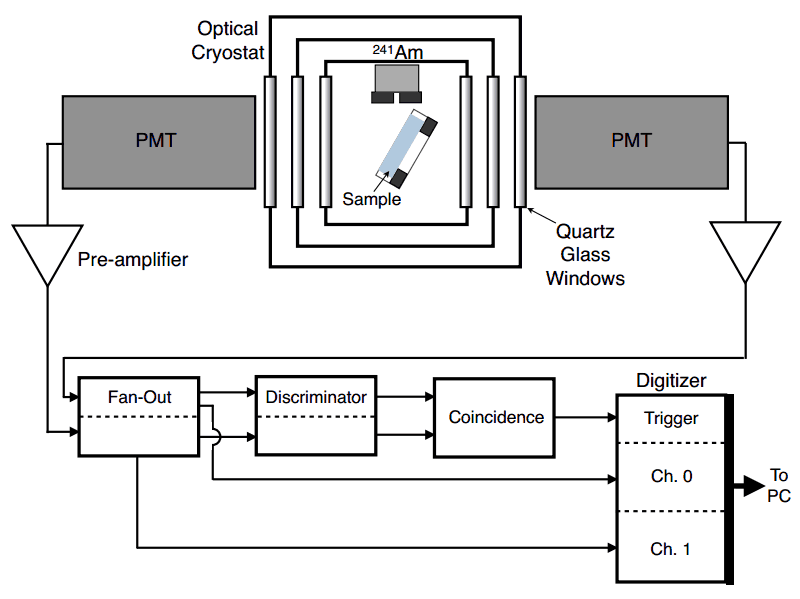}
\setlength{\belowcaptionskip}{10pt plus 3pt minus 2pt}
\caption{\small A schematic of the experimental setup. The sample is placed inside the cryostat under vacuum. The distance between the outside of the outer windows of the vacuum can is 3.2~cm. Within a mm of each window is a PMT viewing the sample.  There is no particular optical coupling between the PMT and the window.  In this figure, the left PMT faces the coated side of the sample (i.e.: it is the strong PMT), while the right PMT faces the blank side.  The angle of incidence of the alphas from the collimated $^{241}$Am source is 35~degrees. Events are triggered if the two PMTs see one or more photelectrons each within a 30~ns coincidence window.}
\label{fig:setup}  
\end{figure*}

\begin{figure*}[!t]
\centering
\subfigure[]{\label{fig:a}\includegraphics[width=70mm]{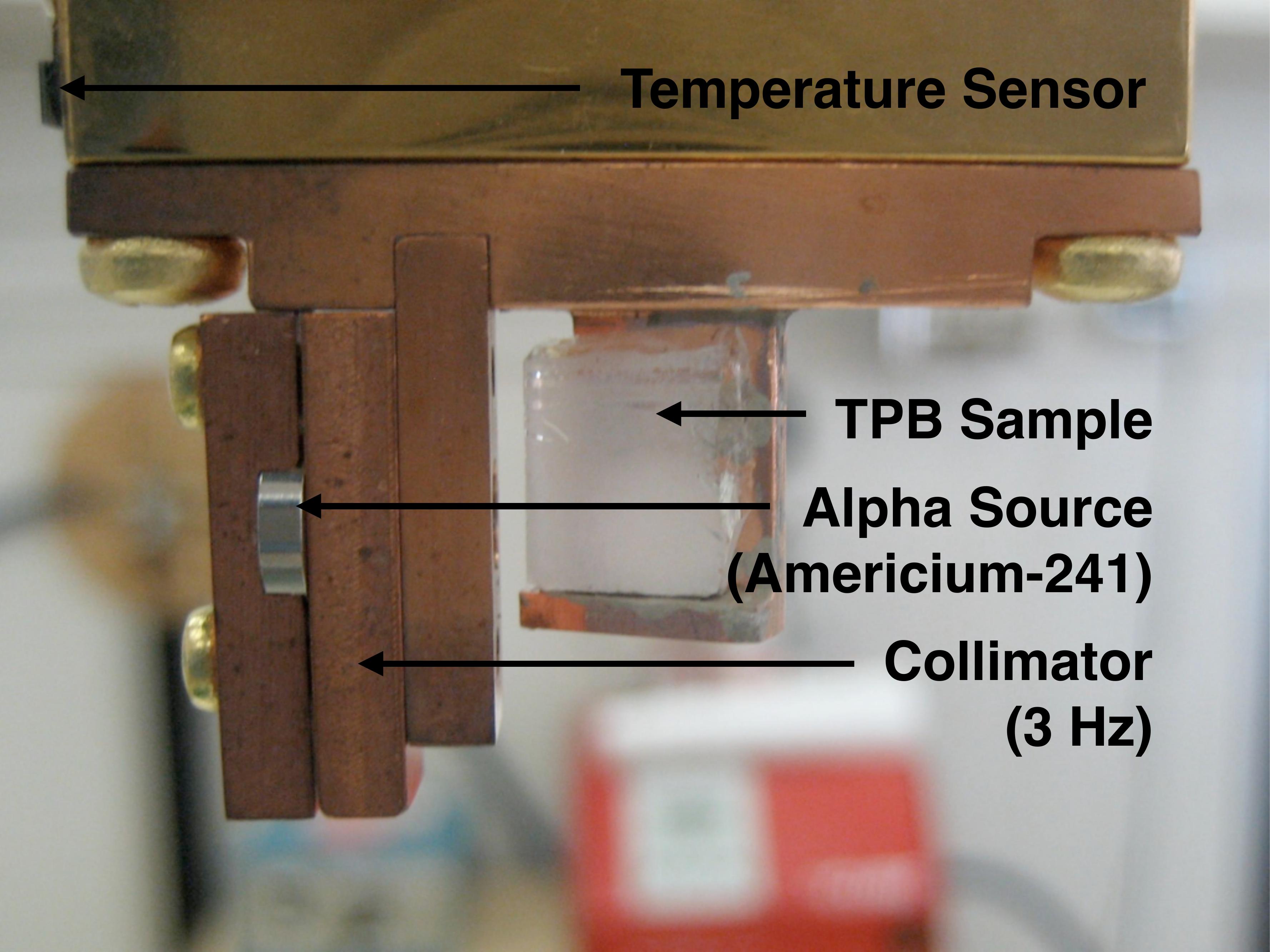}}
\subfigure[]{\label{fig:b}\includegraphics[width=70mm]{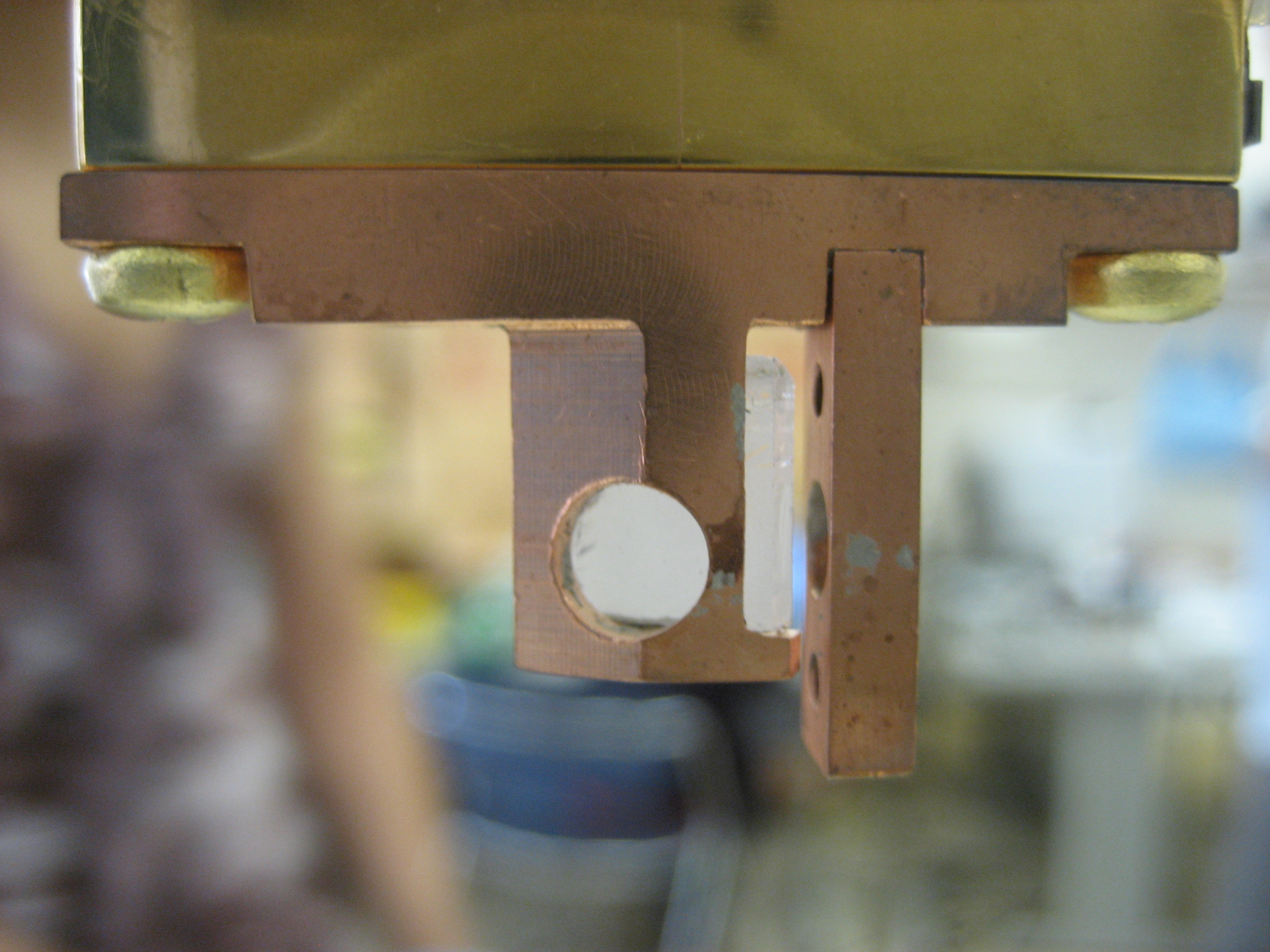}}
\caption{\small Mounted sample with Americium source and 3 Hz collimator installed, front (a) and back (b). Sample is set such that the alphas will hit the smoothest portion of the TPB coating.}
\label{fig:mountedsample}
\label{fig:87Kps}  
\end{figure*}

A TPB sample was prepared using the evaporation technique from~\cite{TinaPaper} (discussed in detail in~\cite{TinaThesis}). For the sample used in this experiment, a 10~\micro\metre\ thick TPB layer was deposited on one face of a quartz (99.995\% silicon dioxide from McMaster-Carr) substrate with dimensions 7$\times$10~mm$^2$ and 1.6~mm thick. The sample was placed in an optical cryostat with compact geometry at Queen's University~\cite{Verdier2009, DiStefano2012}. 
In this device, the bottom tip of the Gifford-McMahon cryocooler is instrumented with a Si diode thermometer and a heater to control the temperature.  A gold-plated oxygen-free high conductivity (OFHC) extension is attached to the tip; a second Si diode measures its temperature in the near vicinity of where the OFHC sample holder screws on to the extension. A Cryo-Con Model 24C temperature controller monitors both thermometers, and uses the heater to regulate the temperature to within 0.1~K.
The sample was centered between two Hamamatsu R6095P photomultiplier tubes (PMTs), which were operated outside the cryostat at room temperature, and viewed the sample through suprasil windows without any particular optical coupling. The distance between the centre of the cryostat and each photomultiplier is $\sim 17$~mm, allowing the PMTs to cover together $\sim 20$\% of the solid angle. The sample was irradiated by a $^{241}$Am alpha source with a thin protective layer degrading the alpha energy to $4.7$~MeV. The source was collimated with a copper aperture to provide 3~Hz of effective event rate (see Figures~\ref{fig:setup} and~\ref{fig:mountedsample}).

As illustrated in Fig.~\ref{fig:setup}, the signals from each PMT were fanned out into a PXI-5154 digitizer module from National Instruments and recorded when in coincidence. Both channels were split at the Fan In/Out unit and half the signal was sent to a discriminator with a 10~mV threshold, corresponding to 85\% single photoelectron detection efficiency. Logical signals from the discriminator were used to generate the coincidence tag if they arrived within the 30~ns coincidence window. The coincidence signal triggered the digitizer to record the PMT signals. The full length of the signal acquisition window was 200~\micro\second, chosen to be sensitive to the long time constants, including a 20~\micro\second\ pre-trigger window, with a 1~ns sampling rate. 

A LabVIEW online analysis program was used for initial data reduction, which included: assigning unique IDs to all events, threshold based identification of individual photons (or inseparable groups of photons) within an event and calculating their charge integrals above the baseline. Then, finally, a list of events with IDs, total charge, and an array of individual photon arrival times and charges were recorded for further offline analysis. 
 
\begin{figure*}[!t]
\centering
\includegraphics[width=120mm,scale=1.5]{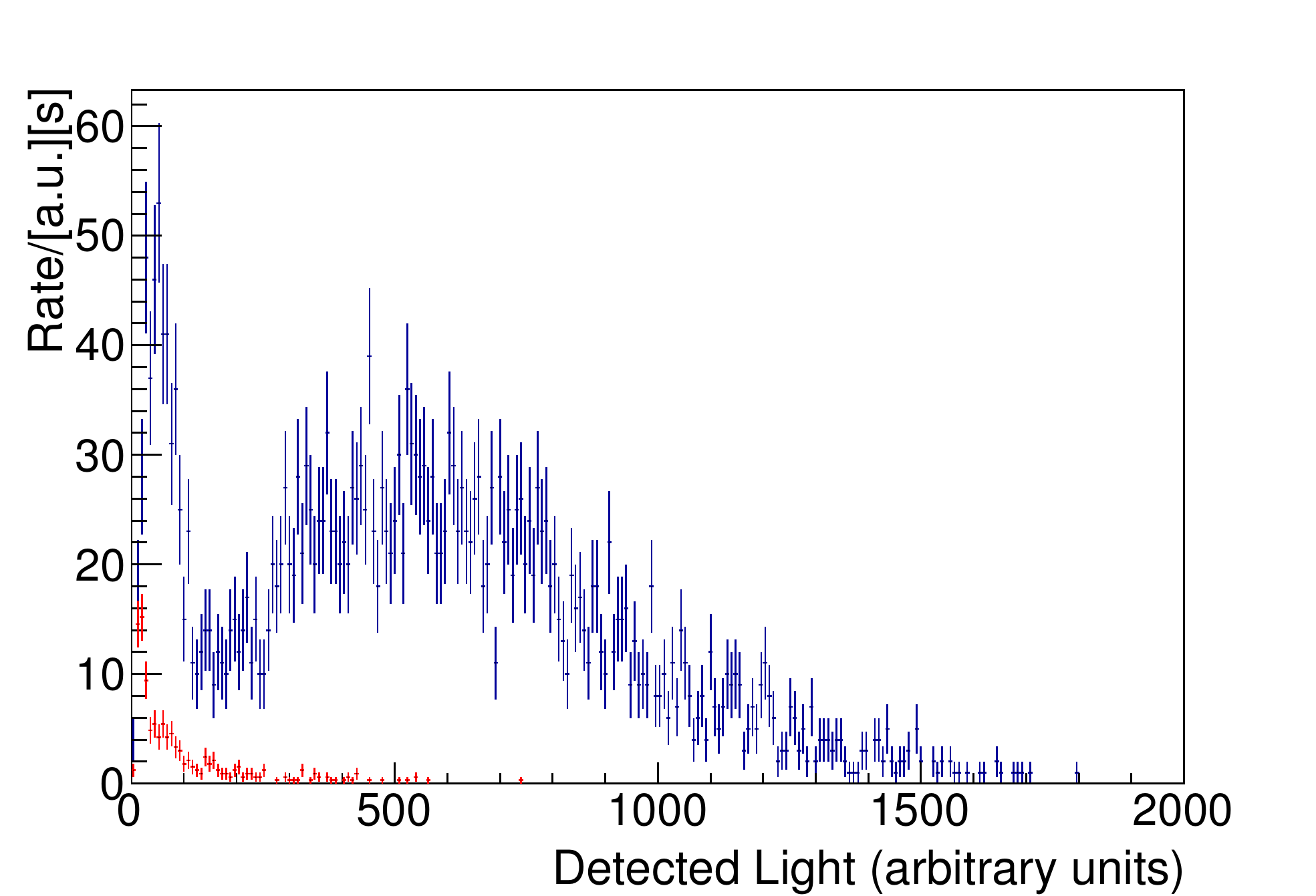} 
\caption{\small Detected light from a blank quartz sample at 87~K (in red) is overlaid onto the detected light from TPB at 87~K, before cuts. Leakage of events from the quartz sample through the data cleaning cuts is negligible.}
\label{fig:quartz}
\label{fig:87Kps}  
\end{figure*}

Scintillation data were recorded at temperatures ranging from room temperature down to 4~K.  After a full temperature sweep, data were taken at room temperature again to verify that no changes had occurred with the sample. In addition, measurements using an identical quartz substrate with no TPB coating were performed to look for any background events that might stem from the substrate rather than the TPB, yielding negative results (see Figure~\ref{fig:quartz}). 

\section{Analysis cuts}\label{sec:cuts}
Two basic data cleaning cuts were applied in the offline analysis in order to remove pile-up events due to two scintillation events arriving within the same time window, pre-trigger events from the long tail of previous events, and random coincidence events~\cite{Kraus2005}. These cuts removed a very small fraction of events (less than 1\%) and involved:
\begin{itemize}
\item The difference in the first photon arrival times between both channels, $\delta t < 5$~ns, to remove pre-trigger events.
\item The first photon arrival time for the weak PMT (PMT1), 19.978~\micro\second$< t_0 <$19.985~\micro\second, also to deal with pre-trigger events.
\end{itemize}

\begin{figure*}[!t]
\center
\subfigure[]{\includegraphics[width=0.5\textwidth]{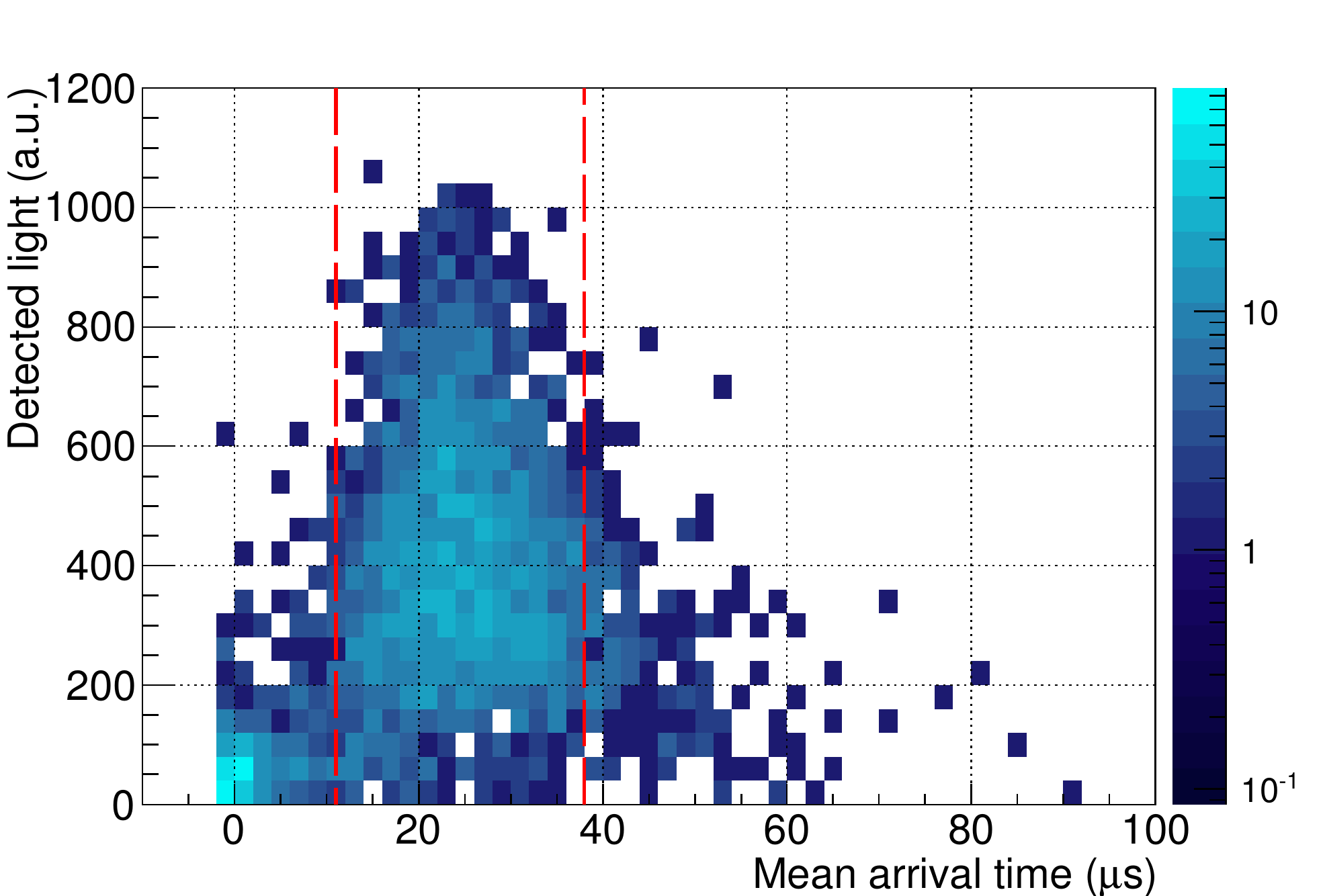}}\subfigure[]{\includegraphics[width=0.5\textwidth]{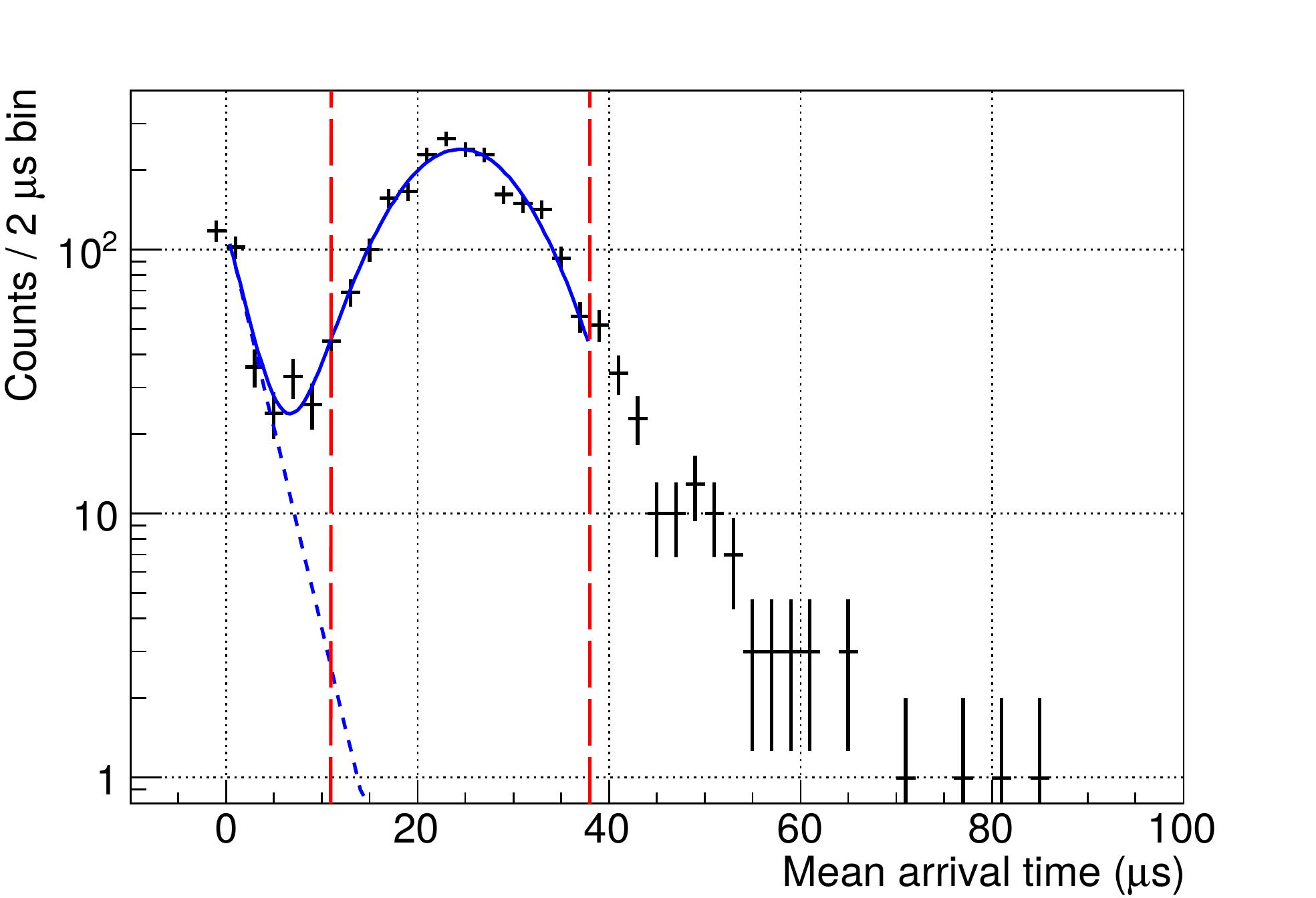}}
\caption{Population of prompt, low charge events is present in the data, here shown at 50~K (see text). (a) 2D histogram in mean arrival time and total event charge. (b) Projection on the mean arrival time axis, with fitted model consisting of a Gaus describing the main population of $\alpha$'s in TPB (blue solid) and a decaying exponential describing the prompt low charge population (blue dashed). On both plots vertical red dashed lines show the position of the mean arrival time cut window. Less than 1\% of events from the cut window come from the prompt low charge population.}\label{fig:meantime}
\end{figure*}

In addition to the population with properties consistent with alphas scintillating in TPB, the data also contained events with relatively low charge and short time constant (see Fig.~\ref{fig:meantime}). 

We have carried out various tests showing that these events correspond to light emission from the TPB.  Use of a thin mask on the source indicates that the events are correlated to the alphas from the source, as opposed to the 60~keV energy gammas from the source.  As fast scintillation of TPB is expected under electron excitation~\cite{Hull2009,Regan94}, one hypothesis is that these events are induced in TPB by low-energy electrons and/or soft X-rays abundantly produced by alphas bombarding the aperture of the copper collimator, as suggested by the literature~\cite{HASSELKAMP1987,Benka95,Incerti2016}.
We have confirmed that the relative contribution of the low charge population was significantly reduced, at a price of increased pile-up, with larger aperture size. We have also observed its gradual disappearance while taking data at increasing pressures of inert and non-scintillating CO$_2$ between 1 and 1000~mbar (while the $\alpha$ population remained unaffected, as expected based on the $\alpha$ range). Both observations support the electron/soft X-ray hypothesis, although the exact composition and spectrum of that secondary emission illuminating the sample, as well as its creation mechanism, remain unclear and are beyond the scope of this paper.

In order to remove that background population from the time component analysis, a cut was applied on the mean event arrival time for the strong PMT (PMT0), computed as charge weighted arrival time average over all pulses in the event (see also \cite{DiStefano2012}). The optimal cut position was manually chosen for each sample temperature. This procedure removed for most temperatures about 20\% of the events (but in some cases up to 35\%).

In terms of distribution of mean arrival times, the low-charge population can be represented by a decaying exponential that accounts for at most 1\% of events passing the cut at each temperature, see Fig.~\ref{fig:meantime}. It has also been cross-checked by taking measurements with a larger collimator aperture and reduced contribution of the background population, which yielded results consistent with those obtained in the standard configuration. The systematic error introduced by the low charge population will be incorporated in Sec.~\ref{sec:syst}.

\section{Results}

\subsection{Detected light and light yield}

\begin{figure}[!t]
\centering
\subfigure[]{\label{fig:b}\includegraphics[width=75mm]{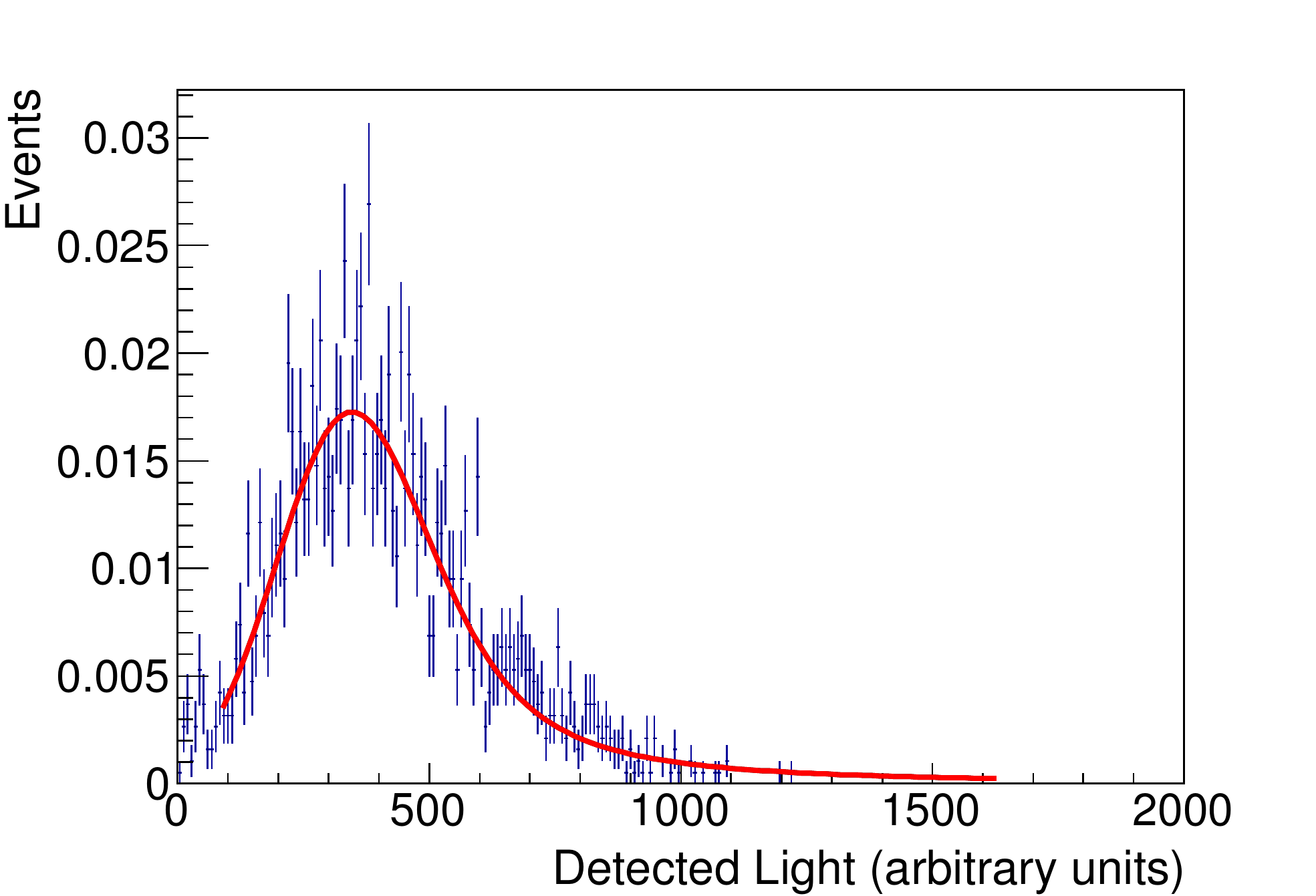}}
\subfigure[]{\label{fig:b}\includegraphics[width=75mm]{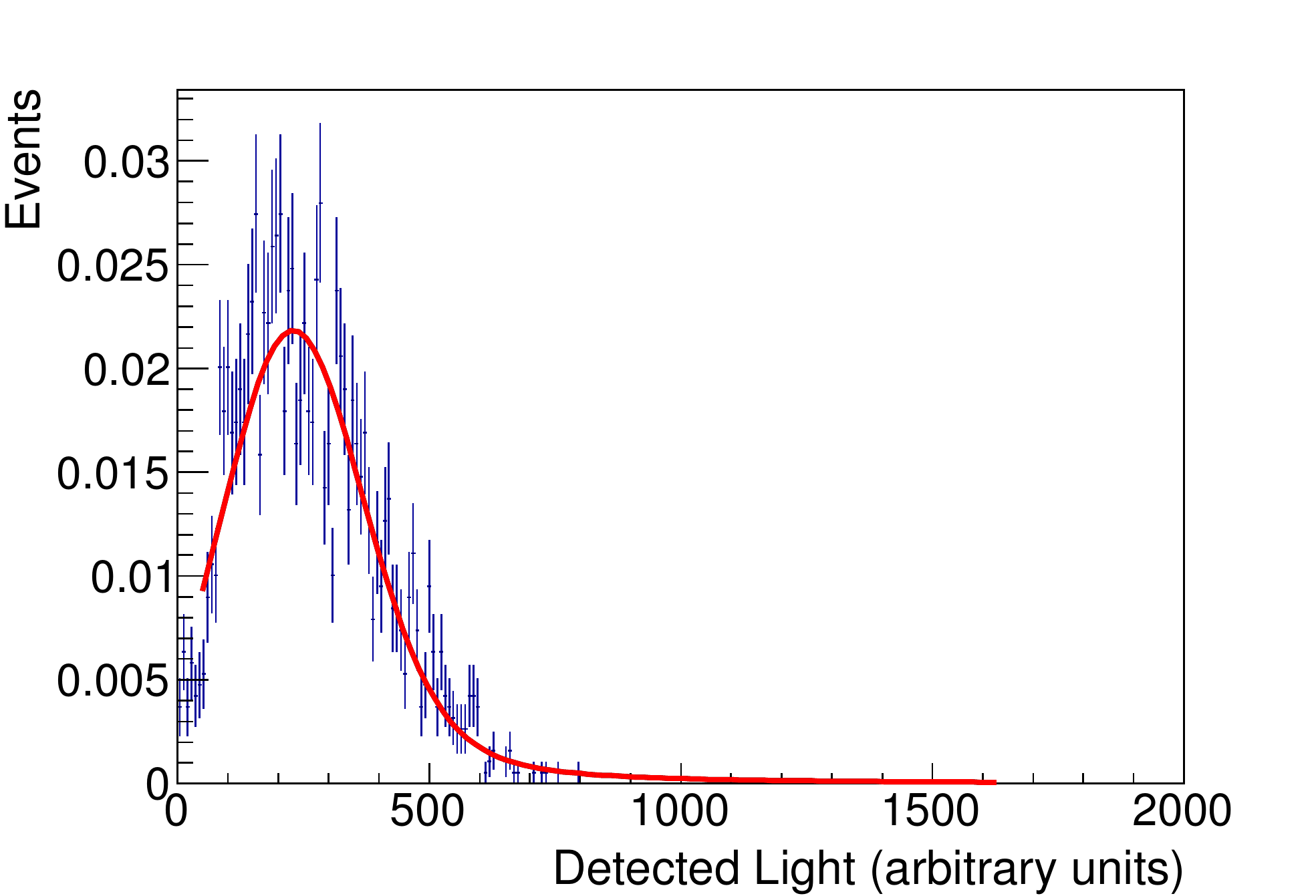}}
\subfigure[]{\label{fig:c}\includegraphics[width=75mm]{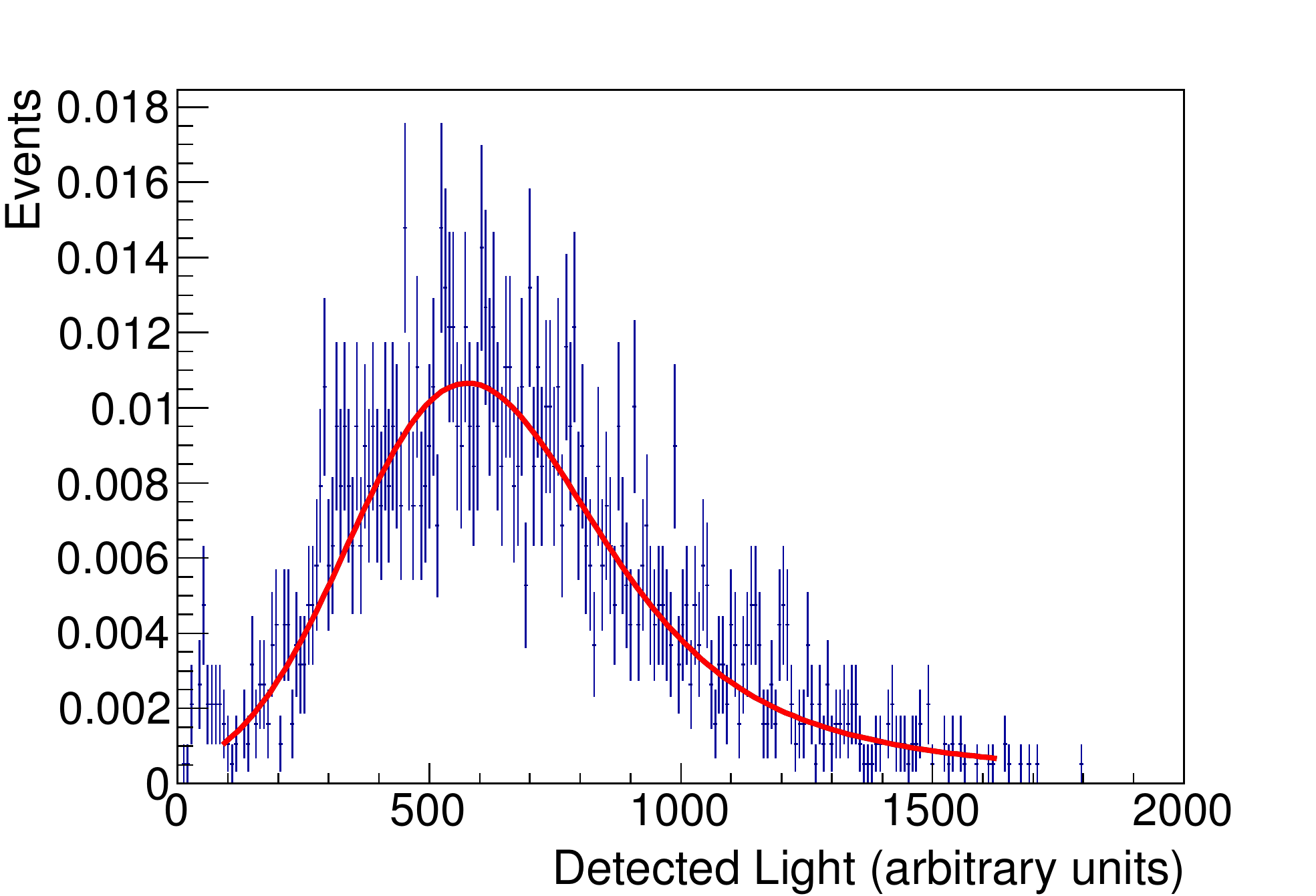}}
\caption[Detected light distribution at 87~K]{Detected light distribution caused by alpha particles interacting in TPB at 87~K, after cuts for (a) PMT0, (b) PMT1, and (c) both PMTs. The area of the histograms is normalized to one.}
\label{fig:87K}  
\end{figure}

\begin{figure*}[!t]
\centering
\subfigure[]{\label{fig:a}\includegraphics[width=70mm]{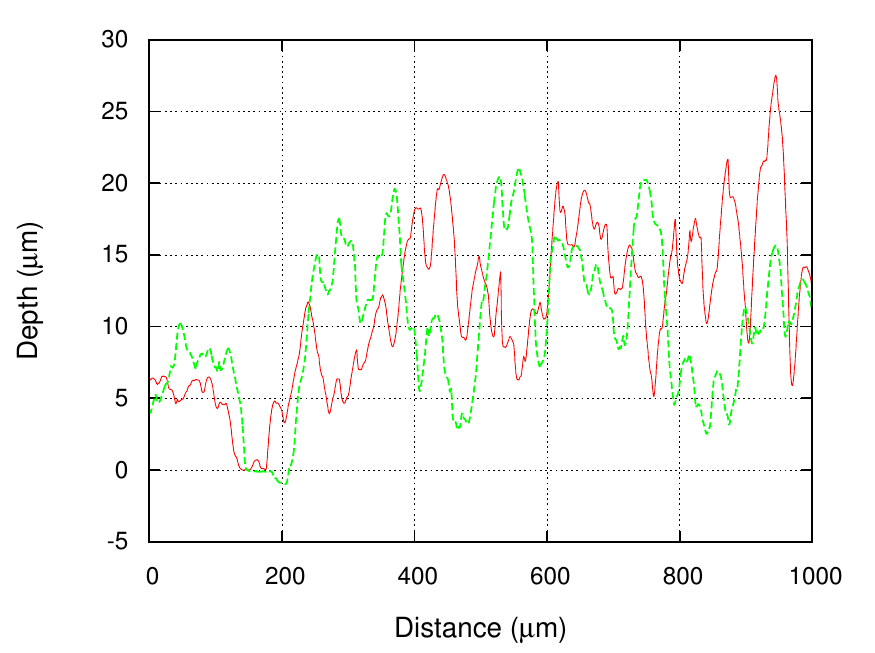}}
\subfigure[]{\label{fig:b}\includegraphics[width=70mm]{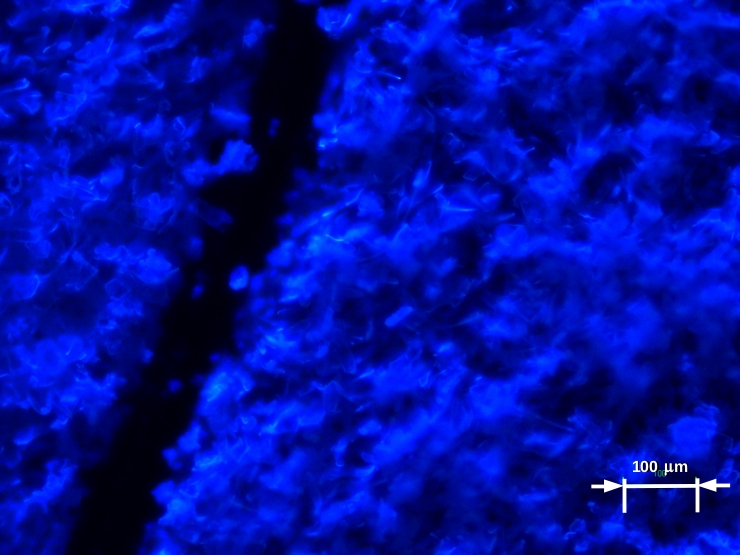}}
\caption{\small (a) TPB thickness was measured using a Dektak 8M Stylus profile meter. The dip between 150~\micro m and 200~\micro m is a scratch, done on purpose to the TPB coating to provide a baseline. The two data sets are from separate scans conducted at different locations on the TPB sample. The TPB coating is rough, with an average thickness of 10~\micro m and 5~\micro m roughness, i.e. the approximate value of roughness parameters $R_a$ and $R_q$. (b) Image of the TPB sample with reference scratch shown, under 355~nm UV illumination. Complex surface morphology is visible, with crystalline structures up to tens of microns in length.}
\label{fig:surface}
\end{figure*}

Figure~\ref{fig:87K} shows the alpha-induced TPB scintillation spectrum at 87~K. The spectrum contains the total light detected by both PMTs. A convolved Landau and Gaussian fitting function was used to find the location of the peaks of the distributions at all temperatures~\cite{PerneggerFriedl2012}. 
A similar fit has been used in other applications, specifically to characterize the energy loss of charged particles through thin layers of silicon~\cite{Meroli2011}. We found that a simple Landau fit could not be used to properly describe the spectra at all temperatures, most likely because the surface roughness of the TPB coating (see Figure~\ref{fig:surface}) and the incident angle distribution of alphas had an additional broadening effect on the distribution. On average, the range of $^{241}$Am $\alpha$ particles in TPB is $\sim$37~\micro m, which is more than the available path length in the coating, i.e. about 17~\micro m (assuming 10~\micro m mean thickness and 35$^\circ$ inclination). Hence for each alpha, the amount of energy deposited in TPB and converted to scintillation light depends on the exact track length in TPB, dominated by the geometry of features in the coating, specific to the exact impact position anywhere within the 0.34~mm$^2$ area beam spot.

\begin{figure*}[!t]
\center
\includegraphics[trim=0cm 0cm 0cm 0cm, clip=true, totalheight=0.4\textheight, angle=0]{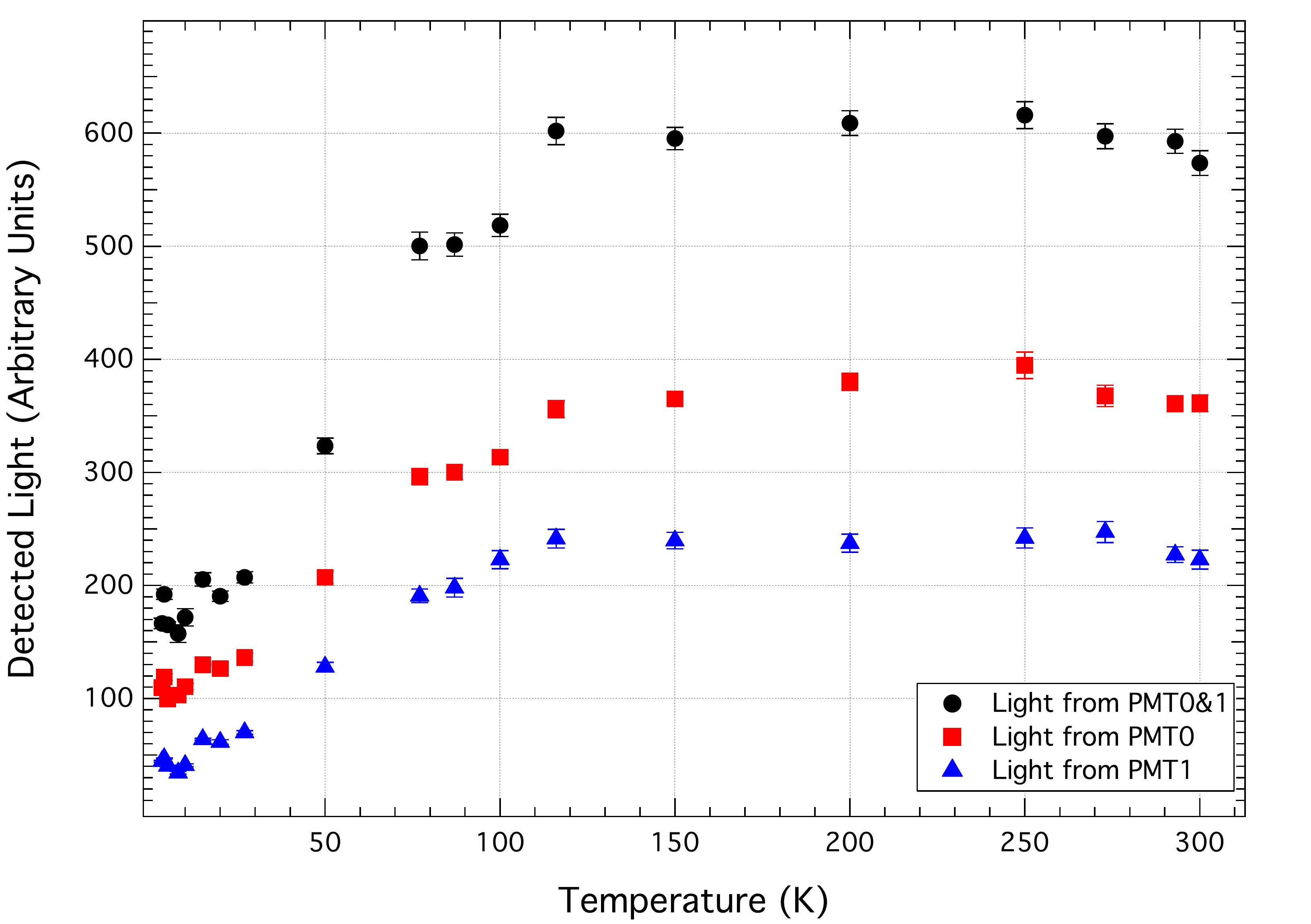}
\caption[Detected light vs. temperature]{Detected light vs. temperature for data between 4~K and 300~K, shown for PMT0, PMT1, and both PMTs}. As the temperature decreases, the light yield also falls for temperatures below 110~K.
\label{fig:LYvsT}  
\end{figure*}

Table~\ref{table:MPVs} lists the most probable values (MPVs) of these distributions at four sample temperatures. The MPVs are related to the peak location of the detected light distributions, and the resulting parameter values from the fits are taken to represent the comparative light yields of TPB at each temperature. Figure~\ref{fig:LYvsT} is a plot of temperature vs. light yield. We can clearly see a reduction in light at temperatures below 110~K in both PMTs, which indicates a reduction in overall light yield of TPB at low temperatures. However, the light yield degrades only marginally at 87~K relative to the room temperature value, which is important for background reduction potential in the DEAP detector.

\begin{table*}[ht]
\center
\begin{tabular*}{0.7\textwidth}{@{\extracolsep{\fill}}  c  c  c  c  c  }
  \hline
  PMT & 300~K & 87~K & 27~K & 4~K \\
   \hline  \hline
  0\&1 & 573$\pm$12 & 500$\pm$10 & 207$\pm$5 & 183$\pm$6\\
   \hline
\end{tabular*}
\caption[Most probable values of light distributions at different temperatures]{Most probable values (MPVs) in arbitrary units for the light yield distribution at four sample temperatures. Errors shown are statistical from the fitting procedure.}
\label{table:MPVs}
\end{table*}

\subsection{Pulse shapes}

The pulse shapes are represented as histograms of the arrival times of the measured photons (or groups of photons), weighted by their integral charge. In order to translate from units of charge to a number of photoelectrons and know the counting statistics needed to set the error bar for each bin, the histograms were divided by the single photoelectron charge (single PE), measured separately. This confirmed that there were adequate statistics for a meaningful fit in the tail of the pulse shape distributions.

Although in general the delayed scintillation of unitary organic scintillators is asymptotically non-exponential~\cite{Laustriat1968}, a multi-exponential model still provides a good match to data and captures the evolution of the main components of the timing spectrum as a function of temperature. The pulse shapes were fit using a convolution of a multi-exponential function with the time resolution function of the measuring system~\cite{Marradanetal2009} with a constant, $c_0$, added to represent noise:
\begin{equation}
F(t) = \left(H(t-t_0) \sum_{i} \frac{N_i}{\tau_i}\,\mathrm{e}^{-(t-t_0)/\tau_i} \right) * G(t) + c_0.
\end{equation}
\noindent Here, $H$ is the Heaviside step function, and the time resolution function is approximated as a Gaussian, $G(t) = \frac{1}{\sqrt{2\pi}\sigma}\mathrm{e}^{-t^2/2\sigma^2}$, where $\sigma$ was determined to be $2 \pm 0.5$~ns (see below). The $\tau_i$ are the time constants, each of respective contribution  $N_i$ to the light yield, and $t_0$ is the start time of the pulse shape.  Four terms were required to satisfactorily fit the data, similar to fits  for alpha-induced scintillation in binary organic scintillators~\cite{Ranuccietal1994}. After convolution the pulse shape $F(t)$ can be written as:
\begin{equation}
\centering
F(t) = \frac{1}{2} \sum_{i=1}^4 \frac{N_i}{\tau_i}\,\mathrm{e}^{\frac{\sigma^2}{2\tau_i^2} - \frac{t-t_0}{\tau_i}}\erfc\left(\frac{\sigma^2 - \tau_i(t-t_0)}{\sqrt{2}\sigma\tau_i}\right) + c_0.
\label{eq:ps}
\end{equation}
The pulse shape fitting procedures implemented here are based on ~\cite{Ranuccietal1994}, \cite{Marradanetal2009} and~\cite{ErinPaper}. After fitting, at each temperature, time constants are normalized to their sum, defining the contribution of each time constant to the light yield as:
\begin{equation}
R_i(T)  \equiv \frac{N_i(T)}{\sum_{j=1}^4 N_j(T)}. 
\label{eq:Ri}
\end{equation}

The first decay time constant, $\tau_1$, has a straightforward relationship to the physics of organic scintillation, as it is very close to the actual value of the lifetime of the lower energy singlet excited state of TPB~\cite{Ranuccietal1994}. The second time constant~\cite{TinaPaper, Kumar1984} can be attributed to further molecular processes such as the de-excitation of triplet spin-states, as is the case for the remaining two long decay constants, which have been first reported for TPB in the course of this work~\cite{LaurelleMSc}, with $\tau_3$ later confirmed in Ref.~\cite{Segreto2015}.

\begin{figure*}[!t]
\centering
\subfigure[]{\label{fig:a}\includegraphics[width=70mm]{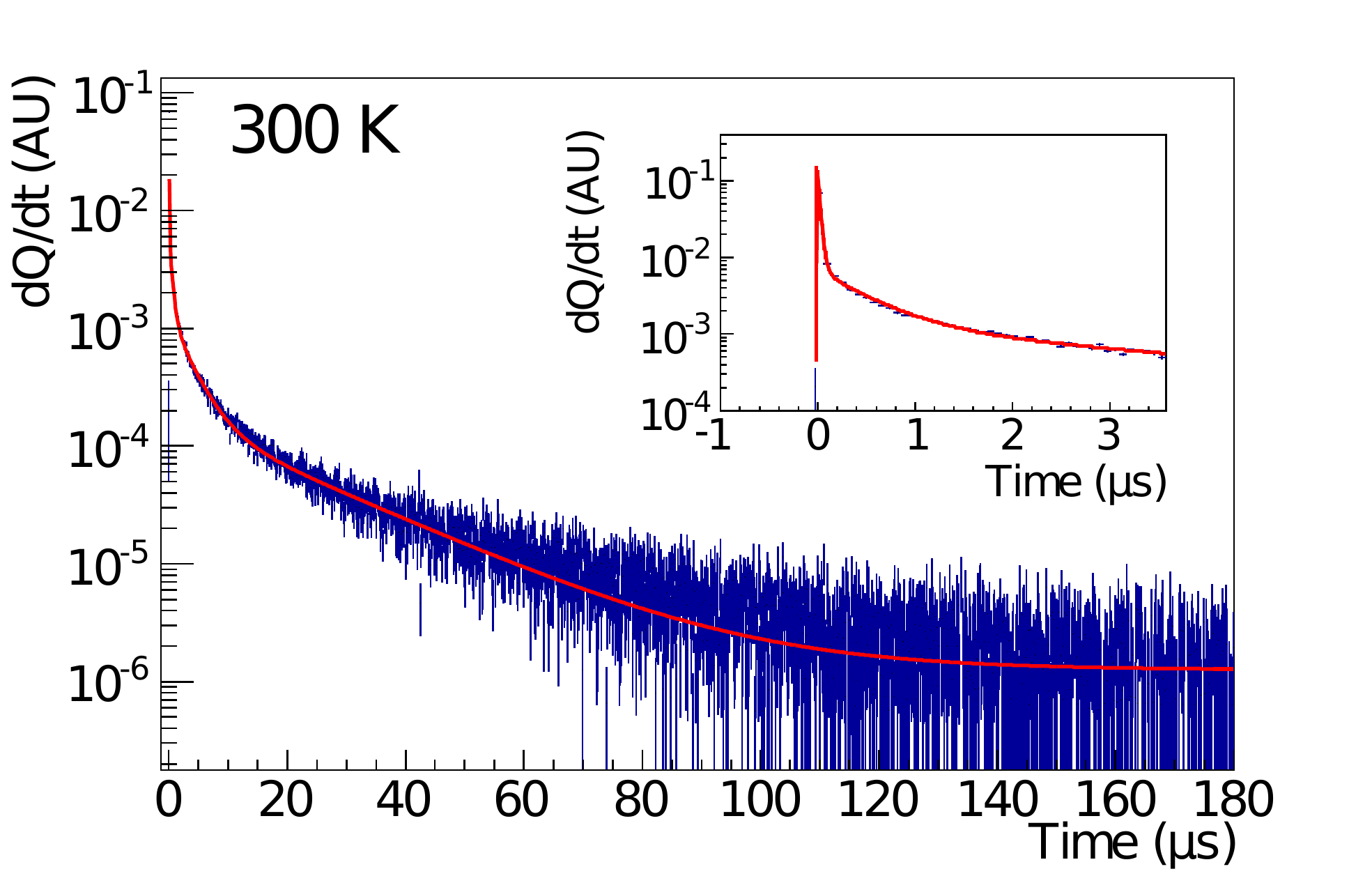}}
\subfigure[]{\label{fig:b}\includegraphics[width=70mm]{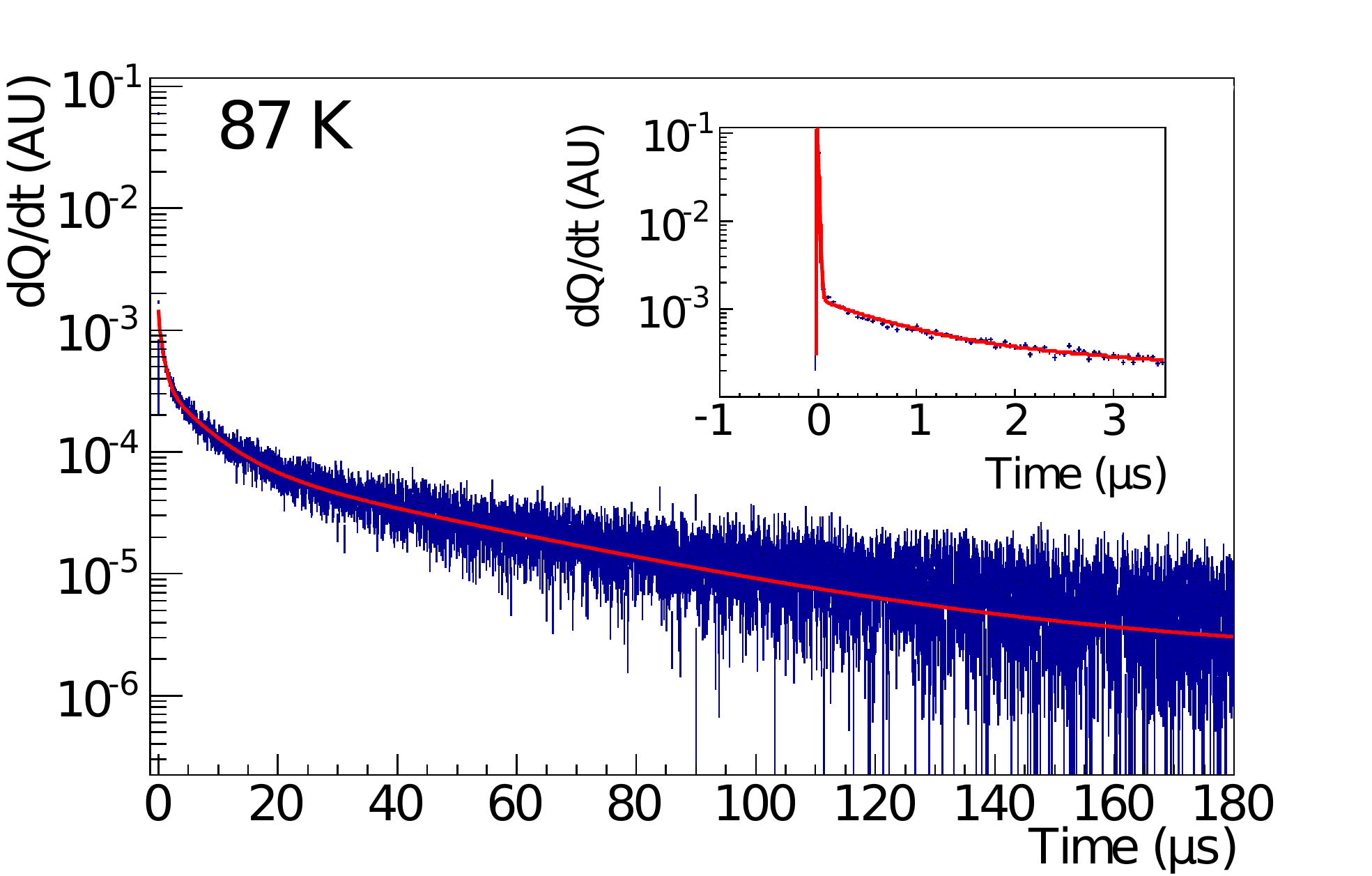}}
\subfigure[]{\label{fig:c}\includegraphics[width=70mm]{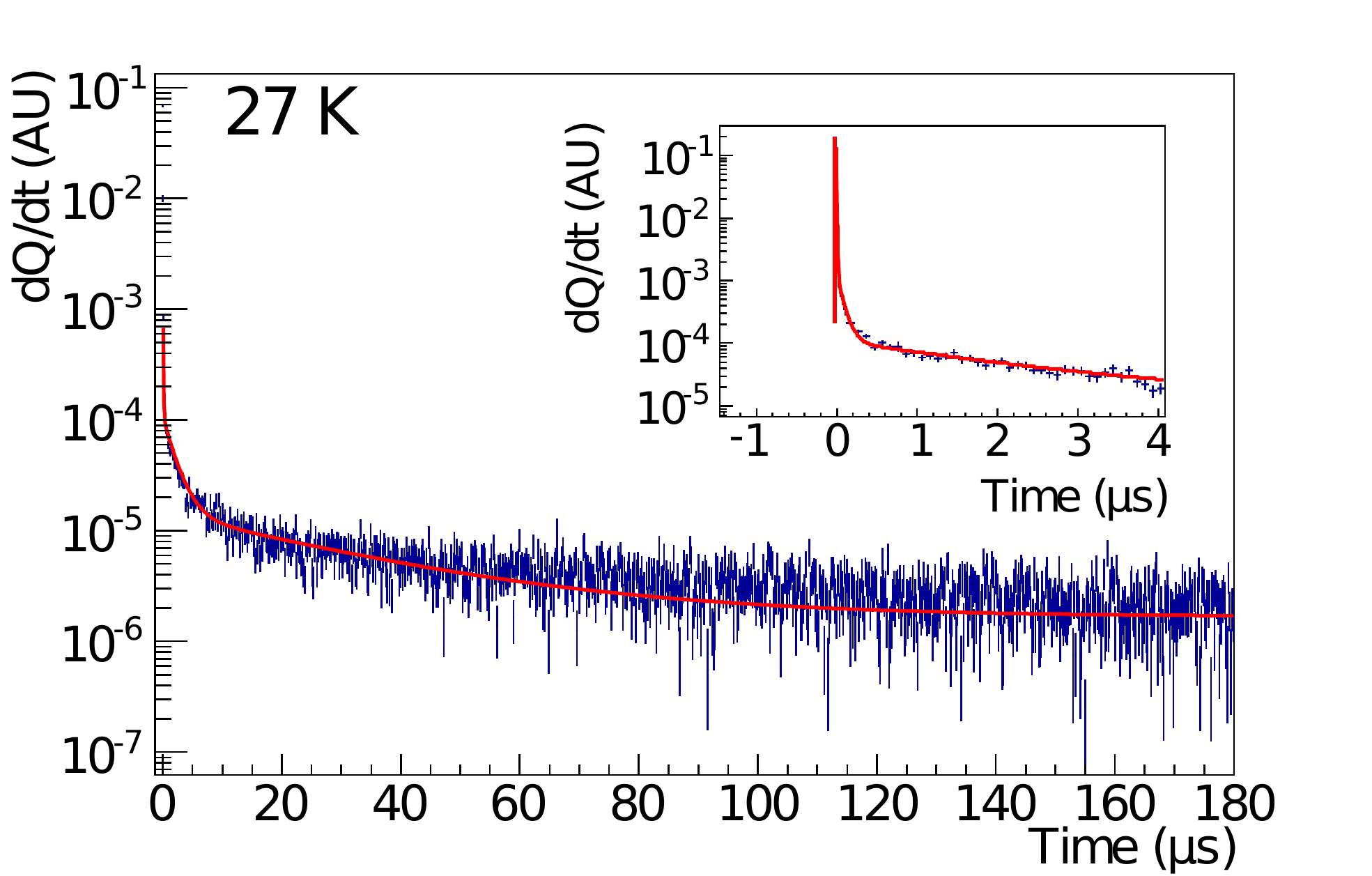}}
\subfigure[]{\label{fig:d}\includegraphics[width=70mm]{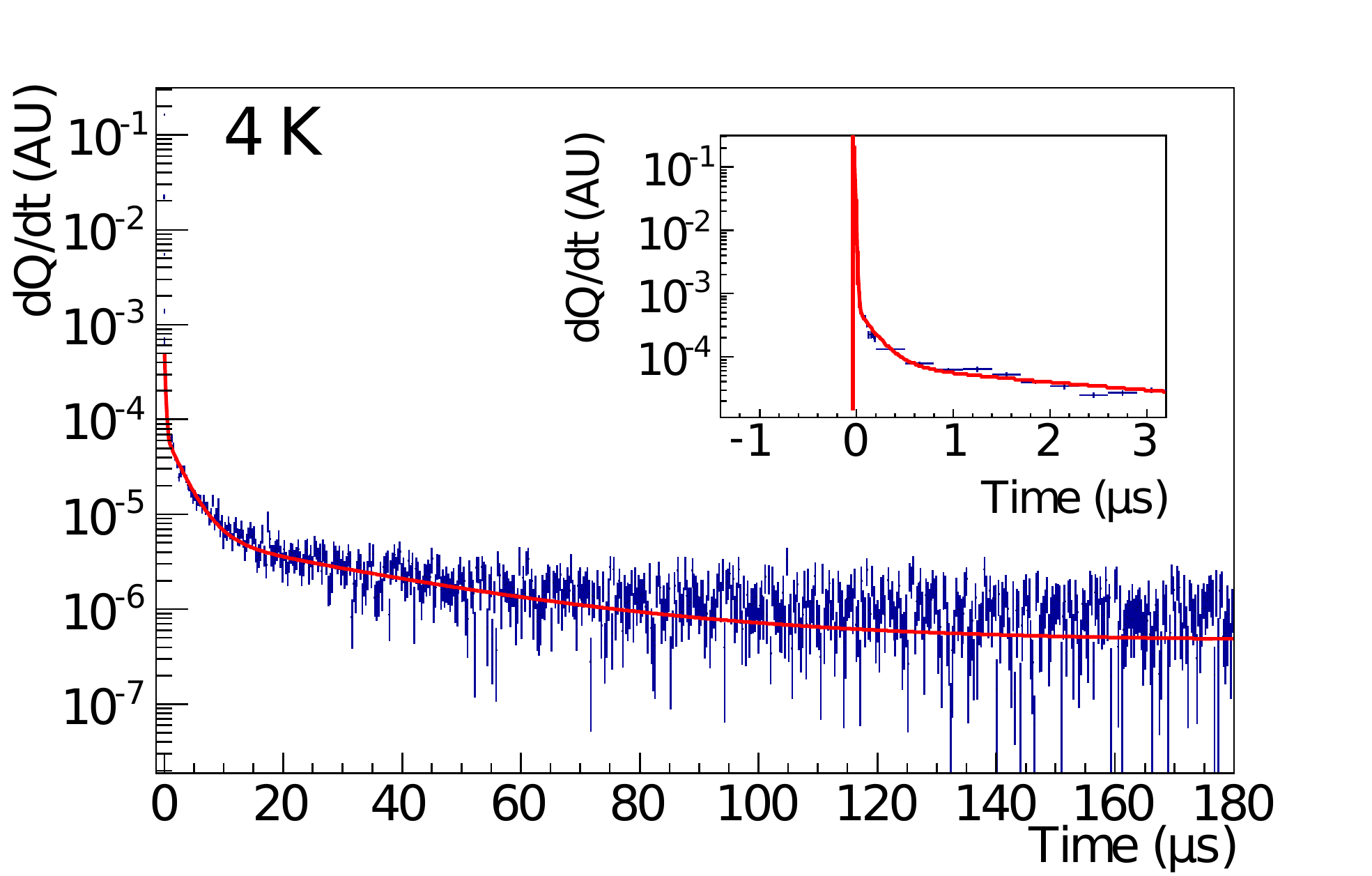}}
\caption[Pulse shape at 87~K]{Examples of pulse shapes, averaged over many events, and fits at various temperatures.  The inserts show the first few microseconds of the pulse.  Pulses are normalized to an area of one. The late light is suppressed as the temperature decreases.}
\label{fig:87Kps}  
\end{figure*}

\begin{figure*}[!t]
\center
\includegraphics[trim=0cm 0cm 0cm 0cm, clip=true, totalheight=0.4\textheight, angle=0]{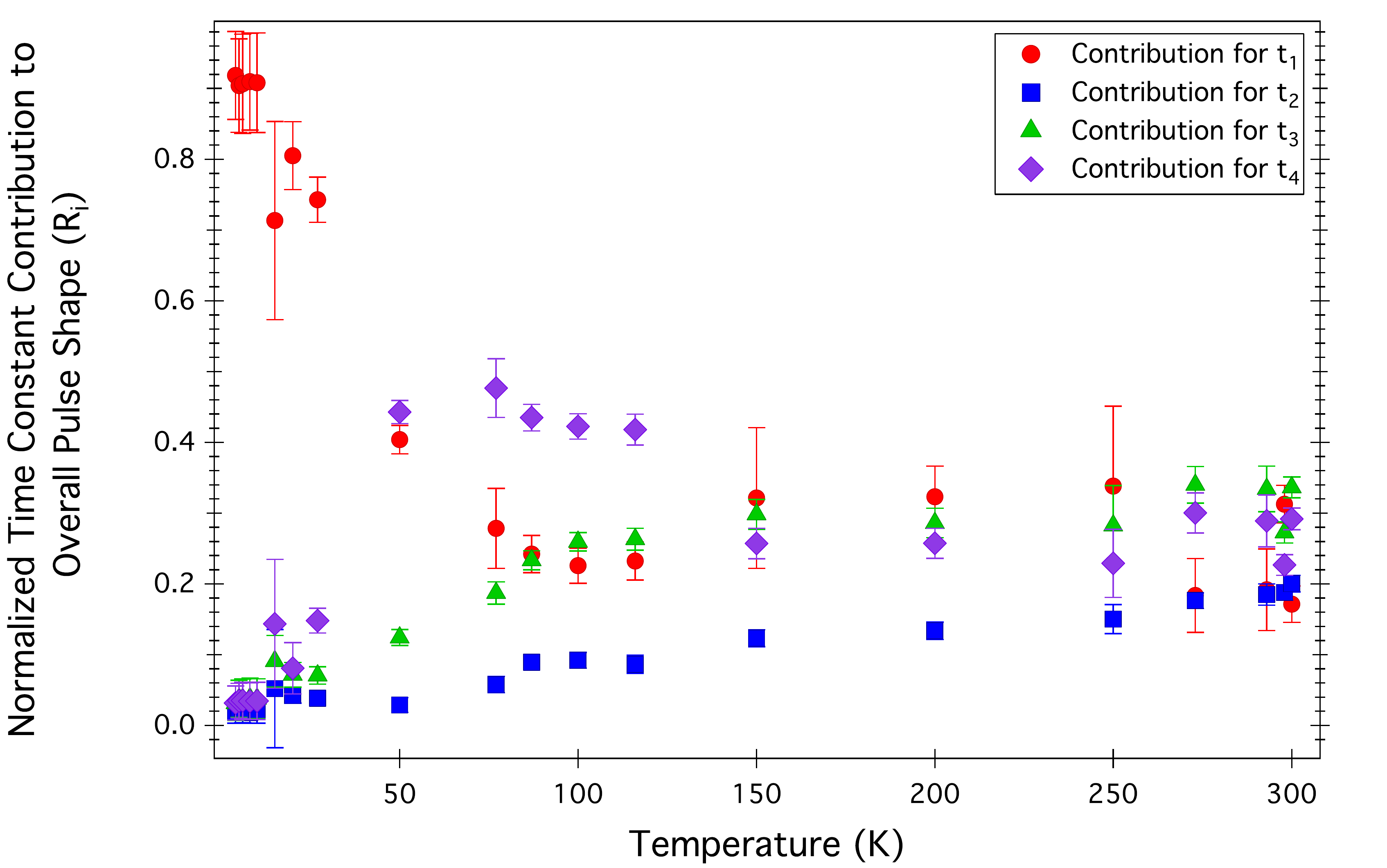}
\caption[Amplitude of time constants vs. temperature]{The contribution of all four time constants  to the light yield, normalized for comparison according to Eq.~\ref{eq:Ri}  ($R_{1,2,3,4}$ for $\tau_1 < \tau_2 ... < \tau_4$). The statistical and systematic errors shown on the plot were added quadratically. }
\label{fig:NvsT}  
\end{figure*}

\begin{figure*}[!t]
\center
\includegraphics[trim=0cm 0cm 0cm 0cm, clip=true, totalheight=0.4\textheight, angle=0]{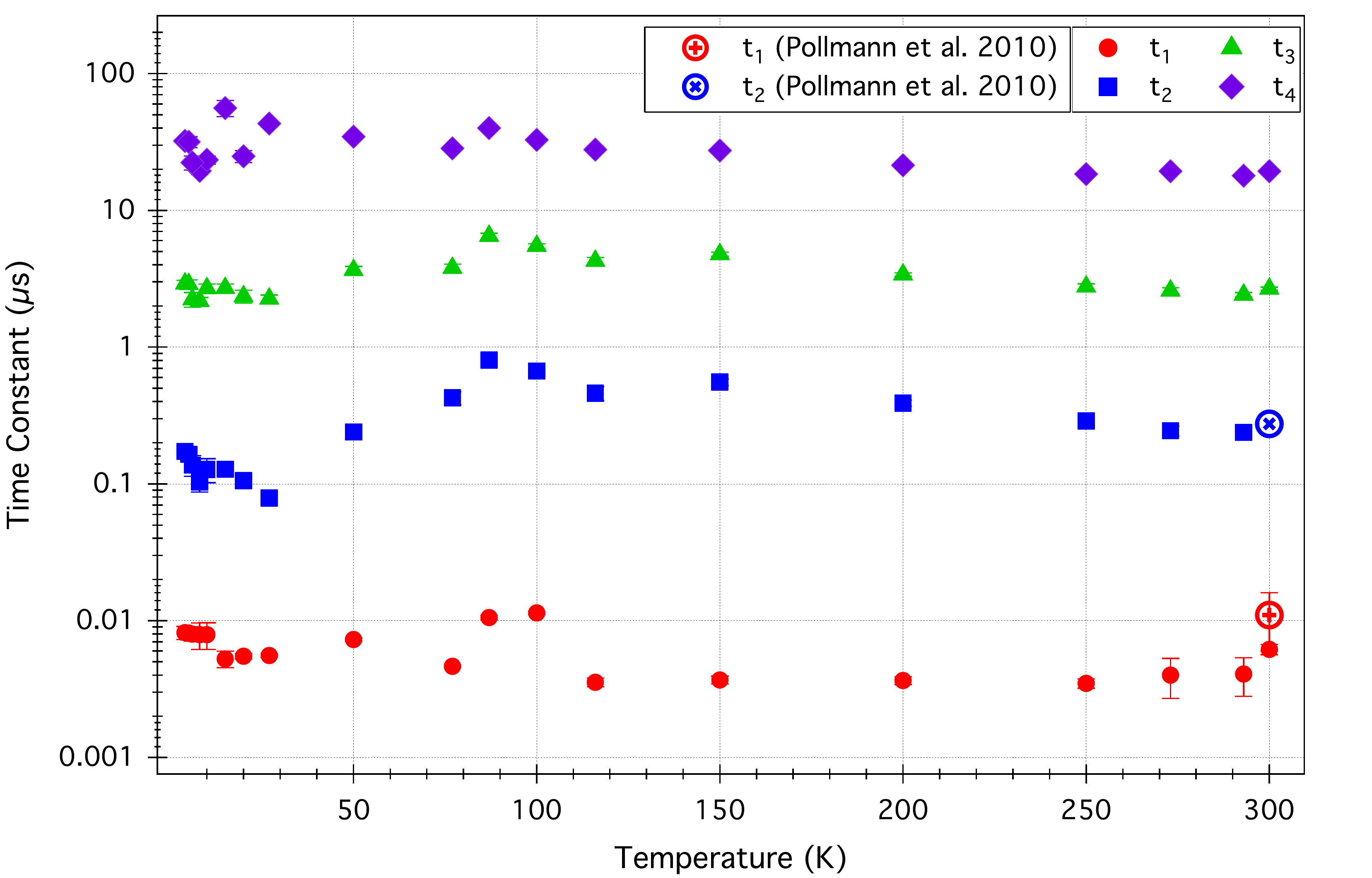}
\caption[Time constants vs. temperature]{Time constants with respect to temperature (300 K - 4 K). Errors include both statistical and systematic uncertainties, added quadratically. The room temperature (RT) measurements from~\cite{TinaPaper} are also shown. Note that these room temperature time constants were fit separately in~\cite{TinaPaper}, and their associated errors were determined by varying the time windows of each fit.}
\label{fig:tvsT}  
\end{figure*}

Figure~\ref{fig:87Kps} shows the pulse shape and resulting fit at four chosen temperatures and demonstrates significant differences between them. Figure~\ref{fig:NvsT} shows that the contribution from $R_1$ dominates the pulse shape at low temperatures, indicating that most of the scintillation light is emitted by the quickly de-exciting singlet energy state. Figure~\ref{fig:tvsT} demonstrates that the time constants $\tau_1$ to $\tau_4$ do not change drastically as the TPB sample is cooled from room temperature. The room temperature measurements from~\cite{TinaPaper} are also shown, with $\tau_2$ consistent with the new result. The first (prompt) time constants are, on average, consistently shorter (though often within systematic uncertainty) than the value found by~\cite{TinaPaper}. This is attributed to systematic uncertainty of the previous measurement (where the time constant was not de-convolved from the time resolution of the PMT).


\begin{table*}[!t]
\center
\small
\begin{tabular*}{1\textwidth}{@{\extracolsep{\fill}}  c|c  c  c  c  }
  \hline
   & 300~K (RT) & 87~K (LAr) & 27~K (LNe) & 4~K (LHe)\\
  \hline \hline
 $\chi^2 /  dof$ & 1.11 & 1.16 & 1.38 & 2.65 \\
 \hline \hline
 R$_1$ & 0.17 & 0.24 & 0.74  & 0.92 \\
 & $\pm$ 0.004 $\pm$ 0.03 & $\pm$ 0.006 $\pm$ 0.03 & $\pm$ 0.04 $\pm$ 0.03 &  $\pm$ 0.02 $\pm$ 0.06 \\
  \hline
    $\tau_1$ & 6.2 & 10.5 & 5.6 & 8.1 \\
  (ns) &  $\pm$ 0.1 $\pm$ 0.5 & $\pm$ 0.3 $\pm$ 0.2 & $\pm$ 0.09 $\pm$ 0.1 & $\pm$ 0.08 $\pm$ 0.9 \\
  \hline \hline
 R$_2$ & 0.200 & 0.089 & 0.038 & 0.02 \\
 & $\pm$ 0.004 $\pm$ 0.01 & $\pm$ 0.004 $\pm$ 0.01 & $\pm$ 0.002 $\pm$ 0.01 & $\pm$ 0.001 $\pm$ 0.02\\
 \hline
  $\tau_2$ & 0.28  & 0.80 & 0.079 & 0.17  \\
 (\micro\second) & $\pm$ 0.005 $\pm$ 0.01 & $\pm$ 0.04 $\pm$ 0.001 & $\pm$ 0.008 $\pm$ 0.001 & $\pm$ 0.01 $\pm$ 0.01 \\
 \hline \hline
 R$_3$ & 0.34 & 0.233 & 0.071 & 0.031 \\
 & $\pm$ 0.004 $\pm$ 0.01 & $\pm$ 0.006 $\pm$ 0.01 & $\pm$ 0.003 $\pm$ 0.01 & $\pm$ 0.002 $\pm$ 0.02 \\
 \hline
  $\tau_3$ & 2.67 & 6.5 & 2.3 & 2.9 \\
 (\micro\second) & $\pm$ 0.07 $\pm$ 0.02  & $\pm$ 0.3 $\pm$ 0.01 & $\pm$ 0.1 $\pm$ 0.02 & $\pm$ 0.2 $\pm$ 0.05 \\
 \hline \hline
 R$_4$ & 0.29 & 0.44 & 0.15  & 0.03 \\ 
 &  $\pm$ 0.004 $\pm$ 0.02 & $\pm$ 0.005 $\pm$ 0.02 & $\pm$ 0.004 $\pm$ 0.02 & $\pm$ 0.002 $\pm$ 0.02 \\
 \hline
  $\tau_4$ & 19.3 & 39.9 & 43.0  & 32.1  \\
 (\micro\second) & $\pm$ 0.3 $\pm$ 0.06 & $\pm$ 1 $\pm$ 0.3 & $\pm$ 2 $\pm$ 0.3 & $\pm$ 3 $\pm$ 0.1 \\
  \hline \hline
\end{tabular*}
\caption{Normalized contribution of the four pulse shape time constants (see Eq.~3.3) and the corresponding time constants at four sample temperatures. Errors shown are statistical and systematic, respectively.}
\label{table:R}
\end{table*}

\subsection{Systematic uncertainties}\label{sec:syst}
The timing resolution, $\sigma$, has a lower limit of 1~ns, determined by the sampling rate of the experiment. The resolution is also affected by additional sources of timing uncertainty. The single photoelectron timing jitter of the PMT (due to photoelectrons being produced at different positions on the cathode), the error in the determination of the start time of the pulse, and the time jitter of the electronic readout chain all influence the achievable resolution of an experiment~\cite{Marradanetal2009}. Because of these factors, we expect the resolution to be wider than the sampling rate. 

To estimate the time resolution, a variety of pulse shapes were fit using different timing resolution values ($\sigma = 1 - 2$~ns). This demonstrated that $\sigma$ only had a significant effect on the shortest time constant ($\tau_1$) and its contribution ($N_1$). In order to cancel that correlation, for the entire temperature sweep we fixed the resolution to a constant value,
which was determined by:
(1) taking the temperature runs with the highest statistics and fitting the entire pulse shape with no fixed parameters, (2) performing fits with all parameters fixed to the results of the first set of fits, except for $N_1$, $\tau_1$, and $\sigma$. As the resulting $\sigma$ values were between 1.5 and 2.5~ns, a value of $\sigma = 2$~ns was fixed for the final analysis, with a systematic uncertainty of 0.5~ns.

Additionally, from a dedicated measurement it was determined that the single photon detection efficiency of the data acquisition was 85\%. The effect of losing approximately 15\% of the single photons was taken into account in the systematic uncertainties by artificially raising the charge of single and double photoelectrons. 

Final systematic errors were estimated by varying the values of $\sigma$ and single PE according to their errors. The variations in the fit parameters were then taken as the systematic errors and added in quadrature, together with 1\% systematic uncertainty introduced by the mean arrival time cut, as discussed in Sec.~\ref{sec:cuts}.

\subsection{Prompt fraction}
 
\begin{figure*}[!t]
\center
\includegraphics[trim=0cm 0cm 0cm 0cm, clip=true, totalheight=0.5\textheight, angle=0]{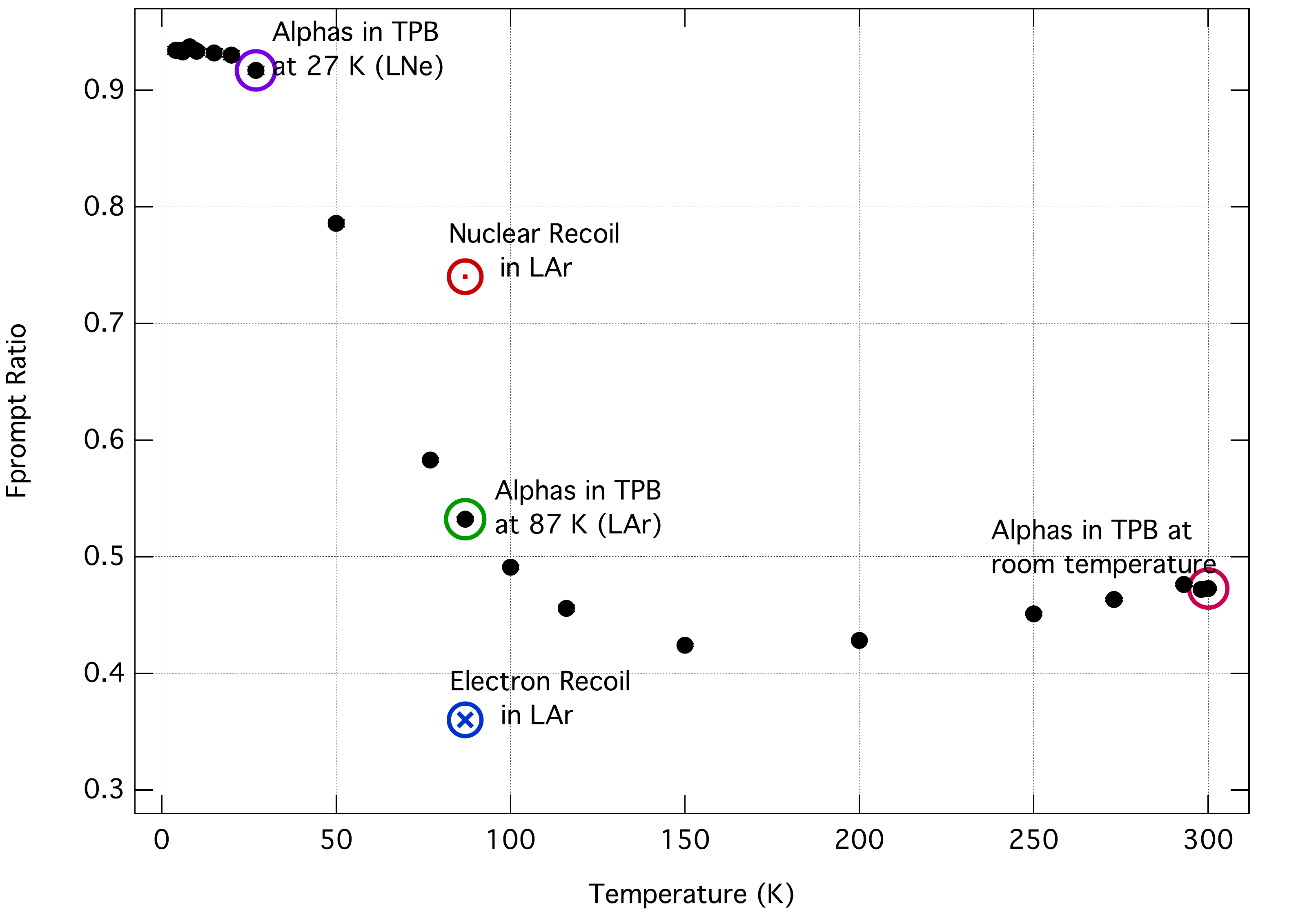}
\caption[\fprompt\ vs. temperature]{\fprompt\ vs. temperature, with important \fprompt\ values labeled. For comparison, alphas in TPB at the temperature of LAr have a ratio of 0.52, nuclear recoils in LAr have a ratio of 0.74, and electron recoils a ratio of 0.36.}
\label{fig:fpromptvsT}  
\end{figure*}

The  prompt fraction, or ``\fprompt\ '', discrimination variable was also investigated using the TPB alpha-induced scintillation measurements. This variable is a measure of the amount of light in the early versus late portions of the pulse shape, where
\begin{equation}
\fpromptmath = \frac{\textrm{PromptPE}}{(\textrm{PromptPE + LatePE})}
\end{equation}
The calculation of \fprompt\ was done using the standard DEAP time windows. PromptPE refers to the pulse shape integral from $t_A - 50$ ns to $t_A + 150$ ns, where $t_A$ is the leading edge of the pulse. LatePE is the integral between $t_A + 150$ ns to 10~\micro\second. The results, with respect to temperature, are shown in Figure~\ref{fig:fpromptvsT}.

In the room temperature measurement published in Ref.~\cite{TinaPaper} a total integration time window of only 1~\micro\second\ was used. This was  due to low statistics and significant level of noise in the late pulse. In this work, by using a more sensitive setup, we found that TPB scintillation at room temperature has a long tail lasting well into the 200~\micro\second\ region, and so used the standard DEAP-1 \fprompt\ time windows (i.e.: a total integration time of 10~\micro\second\ rather than 1~\micro\second), which reduces the \fprompt\ values until about 27~K, when the prompt light begins to dominate.
 Using the short total window from~\cite{TinaPaper} for this analysis, shifts the \fprompt\ value at room temperature from 0.473~$\pm$~0.001 to 0.702~$\pm$~0.002, which is consistent with the room temperature \fprompt\ measurement of 0.67~$\pm$~0.03 from Ref.~\cite{TinaPaper}. Evidently, at temperatures around 27~K and below, the amount of late light is suppressed, which removes the dependence on the choice of total integration window.

\section{Discussion}
\subsection{Temperature dependence of light yield\label{sec:discussion}}
\begin{table*}[!t]
\center
\small
\begin{tabular*}{1\textwidth}{@{\extracolsep{\fill}}  c|c  c  c  c  }
  \hline
Component & 300~K (RT) & 87~K (LAr) & 27~K (LNe) & 4~K (LHe)\\
\hline\hline
1 & 17.0$\pm$3.1 & 20.9$\pm$2.7 & 26.7$\pm$2.0 & 29.4$\pm$2.3 \\
2 & 20.0$\pm$1.2 & 7.8$\pm$1.0 & 1.4$\pm$0.4 & 0.6$\pm$0.7 \\
3 & 34.0$\pm$1.3 & 20.3$\pm$1.1 & 2.6$\pm$0.4 & 1.0$\pm$0.7 \\
4 & 29.0$\pm$2.2 & 38.4$\pm$2.0 & 5.4$\pm$0.8 & 1.0$\pm$0.7 \\
\hline
Sum & 100 & 87.4$\pm$5.2 & 36.1$\pm$2.6 & 32.0$\pm$2.9 \\
\hline
\end{tabular*}
\caption{Percent contribution of each exponential component to overall scintillation yield normalized to the total yield at room temperature ($\frac{N_i(T)}{\sum N_i(\mbox{300~K})}$).}
\label{table:LY}
\end{table*}

Based on Figure~\ref{table:LY} we conclude that the light yields at room temperature and at 87~K agree within 15\%. This is consistent with liquid argon temperature DEAP-1 results~\cite{Amaudruz2015}, which showed agreement between the ex-situ room temperature measurement~\cite{TinaPaper} and a sample of TPB surface alpha decay events extracted from the detector data. However, the time structure of the pulse differs between the two temperatures, most notably due to the long $\sim$40~\micro\second\ component enhanced at 87~K.

We observed a significant decrease in light yield at temperatures below 50~K, which has not been previously reported. At 27~K, we find a light yield that is only 36$\pm 3$\% of the light yield at room temperature and at 4~K it is 32$\pm3$\% (see Table~\ref{table:LY}). In terms of absolute photon yields, slower components 2, 3 and 4 are reduced by a factor of approximately 30 at 4~K with respect to room temperature values. The yield of the fast component is approximately constant between 300 and 87~K and then gradually increases by about 50\% down to 4~K.
For comparison, this is very different than the reported temperature dependence of 128~nm photon induced fluorescence yield of TPB, which exhibits steady $\sim$20\% increase between RT and 12~K~\cite{Francinietal2013, Cavannapresentation}. Complex temperature dependences of $\alpha$- and $\beta$-induced scintillation, different from photon induced fluorescence, have already been reported for organic fluors such as p-terphenyl, anthracene or naphthalene~\cite{Lyons1960,Schiffer1969}.

Possible contributions to the change in the amount of detected light could come from a tem\-perature-dependent shift in the TPB emission peak, as measured in Ref.~\cite{Francinietal2013}, coupled with the efficiency curve of the PMTs used (Hamamatsu R6095P).
We estimate the relative size of this effect to be less than 3\% in the 12-300~K range and with the opposite sign to the trend observed in this work, i.e.  our PMTs become more efficient to the TPB emission spectrum changing with the decreasing temperature. Nevertheless, it is negligible and does not affect any of the conclusions.

 \subsection{Temperature dependence of time components and PSD}
Fig.~\ref{fig:NvsT} shows that the contribution of each time constant to the light yield depends on temperature.
This change can also be seen by the evolution of \fprompt\ with decreasing temperature (Figure~\ref{fig:fpromptvsT}). In fact, by 27~K, most of the light from the longer time constants has been lost; about 90\% of the light is emitted within the first 150~ns of the pulse. This is likely because triplet states are suppressed at lower temperatures. 
Triplet states must be thermally activated in general, since the overlap of vibronic energy levels of S$_1$ and higher order triplet excited states allow intersystem crossing to occur. Evidently, below about 100~K, intersystem crossing becomes less likely, with triplet decays becoming quite rare below around 27~K. This effect is reflected in the observed decrease in light yield, discussed above. It is also evident from Figure~\ref{fig:NvsT} of the normalized time constant amplitudes, that a significant fraction of the total light can be attributed to de-excitations of the longer time constants (i.e.: $\tau_3$ and $\tau_4$). At lower temperatures, these contributions become negligible.

In addition, at 300~K we have confirmed the presence of a 275$\pm$10~ns component to the scintillation pulse shape as measured by~\cite{TinaPaper}, which agrees with our result of $\tau_2$=280$\pm$5 (stat.)$\pm$1 (sys.)~ns. 
Furthermore, we have shown the presence of additional time constant components due to the presence of an unexpectedly long tail in the pulse shape, especially at temperatures above 27~K. These long time constants at room temperature are 2.7$\pm$0.07 (stat.)$\pm$0.02 (sys.)~\micro\second, and 19.3$\pm$0.3 (stat.)$\pm$0.06 (sys.)~\micro\second. We also measure a short time constant at room temperature of 6.2$\pm$0.1~(stat.)$\pm$0.5 (sys.)~ns.

At 87~K, we measure a short time constant of 10.5$\pm$0.3 (stat.)$\pm$0.2 (sys.)~ns, which agrees with the short time constant measured by Segreto~\cite{Segreto2015} for TPB excitation using liquid argon scintillation. We observe an intermediate time constant, $\tau_3$=6.5$\pm$0.3 (stat.)$\pm$0.01 (sys.)~\micro\second, which differs from the intermediate time constant of 3.55$\pm$0.5~\micro\second\ from~\cite{Segreto2015}; we also do not see evidence of the 50~ns decay component reported therein.
We found that the difference in the size of the window used for fitting can explain most of the intermediate time constant discrepancy and constraining the fit to 4.7~\micro\second\ after the peak, as used by Segreto, results in $\tau_3$=4.7$\pm$0.4, which is only marginally inconsistent with his value.
Similarly, our longest decay constant, which exceeded 20~\micro\second, has not been reported before, possibly because the acquisition window size in the published results is much shorter than 200 200~\micro\second\  used in our measurement. 

\begin{figure*}[!t]
\center
\includegraphics[trim=0 0 12cm 20cm, clip=true, totalheight=0.5\textheight]{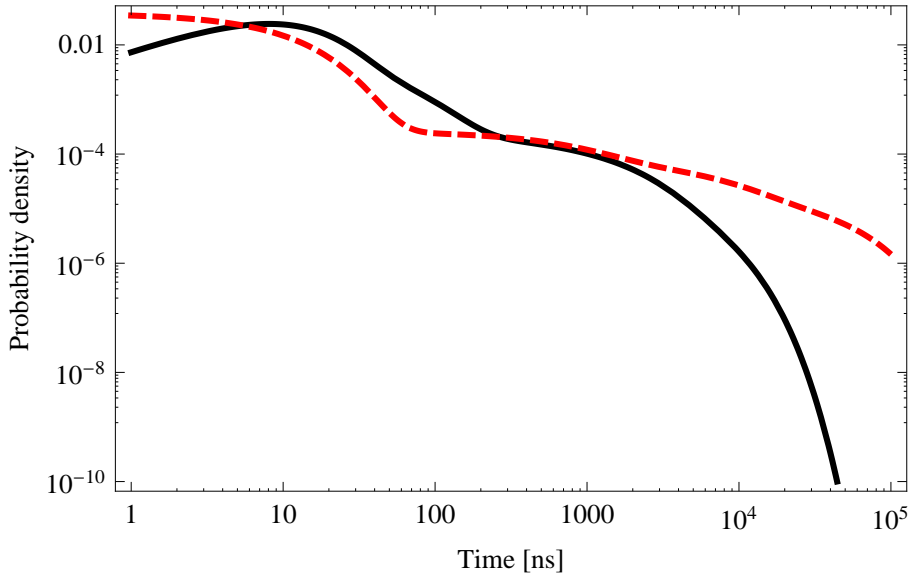}
\caption{Photon emission time probability density function for TPB photons induced by alpha excitation at 87~K based on this work (red, dashed) and excitation by 128~nm photons from nuclear recoil scintillation in liquid argon~\cite{Segreto2015} (black).}
\label{fig:TpbAr}  
\end{figure*}

Our measured \fprompt\ values improve on the measurement done by~\cite{TinaPaper}. The \fprompt\ of alpha-induced TPB scintillation at liquid argon temperatures is 0.512$\pm$0.002, whereas the \fprompt\ of DEAP-1 nuclear recoils in the liquid argon itself is 0.8.
Using the alpha scintillation pulse shape at 87~K from this work we can now construct the optimal parameter for discriminating surface alpha decay events from genuine nuclear recoil events in liquid argon. Figure~\ref{fig:TpbAr} compares pulse shapes for both types of events, where the nuclear recoil pulse shape was constructed as a convolution of a standard Ar scintillation model with TPB response function from Ref.~\cite{Segreto2015}.

Although more elaborate PSD methods (e.g. based on a likelihood ratio) might be more effective, in order to show a simple but informative example we will use the charge fraction method and construct an equivalent of the \fprompt\ parameter.
The charge fraction which maximizes the separation between nuclear recoil and alpha probability density functions relies on the time window constrained by the intersection points between both PDFs in Figure~\ref{fig:TpbAr}, i.e. $t_1$=6~ns and $t_2$=270~ns,
\begin{equation}
F_\alpha \equiv \frac{\int_{t_1}^{t_2}I(t)dt}{\int_0^{t_{max}}I(t)dt},
\end{equation}
where $I(t)$ is the intensity of the measured scintillation signal, and $t_{max}$ is the integration window.
Using $t_{max}$ = 10 \ \micro\second\ (practical for discriminating electronic recoils from nuclear recoils in LAr) results in a difference in $F_{\alpha}$ between alphas and nuclear recoils of $\Delta^\alpha$=0.35 (compared to 0.4 separation in \fprompt\ between electronic and nuclear recoils). The separation $\Delta^\alpha$ can be increased to 0.4 with t$_{max}$=30~\micro\second, which allows to take additional advantage of the delayed TPB emission. It remains an open question whether the corresponding longest time constant exists in VUV florescence light (beyond the time window studied in Ref.~\cite{Segreto2015}) or not, which is assumed here.  

One radioactive background relevant for dark matter searches is when a nucleus (for instance a $^{222}$Rn daughter) near the surface of the TPB decays, sending an $\alpha$ particle into the TPB and a nucleus into the LAr. 
Without any way to tag the $\alpha$, the nucleus could mimick a WIMP signal in the LAr, in particular if it looses some energy before entering the LAr. Thanks to the presence of TPB however, these events will contain  photons contributed both from the $\alpha$ in TPB and from the nuclear recoil  in liquid argon, making them identifiable.  In practice,  the actual discrimination factor will depend on the amount of $\alpha$ energy deposited in TPB, hence on the event topology and the TPB coating thickness, which calls for detector-specific simulation studies.
Because of the much smaller contribution of delayed emission from $\alpha$ scintillation at temperatures below $\sim$50~K, it appears that a similar discrimination technique in a liquid neon or liquid helium detector would be  more challenging.

\section{Conclusions}
1,1,4,4-Tetraphenyl-1,3-butadiene is commonly used as a wavelength shifter in liquid noble gas experiments, including dark matter searches. 
We have characterized the light yield and decay times of  TPB excited with $\alpha$-particles to temperatures down to 4~K, and have confirmed the presence of a long scintillation time constant, while also demonstrating the presence of additional long time constants in the late pulse region.
We have observed a significant decrease in light yield at temperatures below 50~K, due to suppressed delayed emission from triplet states, which has not been previously reported for TPB.

From the standpoint of background identification, we have constructed a discrimination parameter optimized for distinguishing TPB-surface alpha decay events from true recoil events in liquid argon. 
TPB also provides an avenue to tag nuclear recoils in LAr caused by an $\alpha$ decay in TPB.  The scintillation properties of TPB may therefore provide an extra background-reduction tool in rare-event searches.

\acknowledgments
Authors are grateful to Gregory Jerkiewicz for granting access to the surface profiler as well as thank Kevin Robbie and Chelsea Elliot for access and assistance with the fluorescence microscope.
Useful comments to the manuscript from Chris Jillings and Fabrice Reti\`ere are grate\-fully acknow\-ledged.
This work has been funded in Canada by NSERC (Grant SAPIN 386432), CFI-LOF and ORF-SIF (project 24536).

\bibliographystyle{JHEP}
\bibliography{tpb}

\providecommand{\href}[2]{#2}\begingroup\raggedright\begin{thebibliography}{10}

\bibitem{deap_proc}
P.-A. Amaudruz et~al. (DEAP), {\it {DEAP-3600 Dark Matter Search}},  {\em
  Nuclear and Particle Physics Proceedings,} {\bf
  \href{http://dx.doi.org/10.1016/j.nuclphysbps.2015.09.048}{doi:10.1016/j.nuclphysbps.2015.09.048}}
  (2016) [\href{http://arxiv.org/abs/1410.7673}{{\tt arXiv:1410.7673}}].

\bibitem{miniclean}
K.~Rielage and {MiniCLEAN Collaboration}, {\it Status and prospects of the
  {MiniCLEAN} dark matter experiment},  {\em AIP Conference Proceedings} {\bf
  1441} (2012) 518--520.

\bibitem{darkside}
T.~Alexander et~al., {\it {DarkSide} search for dark matter},  {\em
  \jinst{8}{2013}{C11021}} (2013).

\bibitem{ardm}
A.~Rubbia, {\it {ArDM}: a ton-scale liquid {Argon} experiment for direct
  detection of dark matter in the universe},  {\em Journal of Physics:
  Conference Series} {\bf 39} (2006) 129.

\bibitem{warp}
P.~Benetti et~al., {\it First results from a dark matter search with liquid
  argon at 87~{K} in the {Gran Sasso} underground laboratory},  {\em
  Astroparticle Physics} {\bf 28} (2008) 495 -- 507.

\bibitem{nedm}
T.~M. Ito and the {nEdm}~Collaboration, {\it Plans for a neutron {EDM}
  experiment at {SNS}},  {\em Journal of Physics: Conference Series} {\bf 69}
  (2007) 012037.

\bibitem{microboone}
T.~Katori, {\it The {MicroBooNE} light collection system},  {\em
  \jinst{8}{2013}{C10011}} (2013).

\bibitem{TinaPaper}
T.~Pollmann, M.~Boulay, and M.~Ku\'zniak, {\it Scintillation of thin
  tetraphenyl butadiene films under alpha particle excitation},  {\em Nucl.
  Instr. Meth. Phys. Res. A} {\bf 635} (2011) 127.

\bibitem{hitachi}
A.~Hitachi et~al., {\it Effect of ionization density on the time dependence of
  luminescence from liquid argon and xenon},  {\em Phys. Rev. B} {\bf 27}
  (1983) 5279--5285.

\bibitem{Boulay2009}
M.~G. {Boulay et al. (DEAP)}, {\it Measurement of the scintillation time
  spectra and pulse-shape discrimination of low-energy $\beta$ and nuclear
  recoils in liquid argon with {DEAP-1}},
  \href{http://arxiv.org/abs/0904.2930}{{\tt arXiv:0904.2930}}.

\bibitem{Segreto2015}
E.~Segreto, {\it Evidence of delayed light emission of tetraphenyl-butadiene
  excited by liquid-argon scintillation light},  {\em Phys. Rev. C} {\bf 91}
  (2015) 035503.

\bibitem{Baker2008}
J.~Baker, N.~Galunov, and O.~Tarasenko, {\it Variation of scintillation light
  yield of organic crystalline solids for different temperatures},  {\em IEEE
  Transactions on Nuclear Science} {\bf 55} (2008) 2736--2738.

\bibitem{TinaThesis}
T.~Pollmann, {\em {Alpha Backgrounds in the {DEAP} Dark Matter Search
  Experiment}}.
\newblock PhD thesis, Queen's University, 2012.

\bibitem{Verdier2009}
M.-A. Verdier et~al., {\it A 2.8~{K} cryogen-free cryostat with compact optical
  geometry for multiple photon counting},  {\em Review of Scientific
  Instruments} {\bf 80} (2009) 046105.

\bibitem{DiStefano2012}
P.~C.~F. Di~Stefano, P.~Nadeau, C.~Onderwater, C.~Trudeau, and M.-A. Verdier,
  {\it Counting photons at low temperature with a streaming time-to-digital
  converter},  {\em Nucl. Instr. Meth. Phys. Res. A} {\bf 700} (2013) 40 -- 52.

\bibitem{Kraus2005}
H.~Kraus, V.~Mikhailik, and D.~Wahl, {\it Multiple photon counting coincidence
  ({MPCC}) technique for scintillator characterisation and its application to
  studies of {CaWO}${}_4$ and {ZnWO}${}_4$ scintillators},  {\em Nucl. Instr.
  Meth. Phys. Res. A} {\bf 553} (2005) 522 -- 534.

\bibitem{Hull2009}
G.~Hull, N.~P. Zaitseva, N.~J. Cherepy, J.~R. Newby, W.~Stoeffl, and S.~A.
  Payne, {\it New organic crystals for pulse shape discrimination},  {\em IEEE
  Transactions on Nuclear Science} {\bf 56} (2009) 899--903.

\bibitem{Regan94}
S.~P. Regan, L.~K. Huang, M.~J. May, H.~W. Moos, D.~Stutman, S.~Kovnovich, and
  M.~Finkenthal, {\it {Measured conversion efficiencies of P45, paraterphenyl,
  tetraphenyl butadiene, and sodium salicylate phosphors in the soft-x-ray
  wavelength range}},  {\em Appl. Opt.} {\bf 33} (1994) 3595--3599.

\bibitem{HASSELKAMP1987}
D.~Hasselkamp, S.~Hippler, and A.~Scharmann, {\it Ion-induced secondary
  electron spectra from clean metal surfaces},  {\em Nucl. Instr. Meth. Phys.
  Res. B} {\bf 18} (1986) 561 -- 565.

\bibitem{Benka95}
O.~Benka, A.~Schinner, T.~Fink, and M.~Pfaffenlehner, {\it {Electron-emission
  yield of Al, Cu, and Au for the impact of swift bare light ions}},  {\em
  Phys. Rev. A} {\bf 52} (1995) 3959--3965.

\bibitem{Incerti2016}
S.~Incerti, B.~Suerfu, J.~Xu, V.~Ivantchenko, A.~Mantero, J.~Brown, M.~Bernal,
  Z.~Francis, M.~Karamitros, and H.~Tran, {\it Simulation of {Auger} electron
  emission from nanometer-size gold targets using the geant4 monte carlo
  simulation toolkit},  {\em Nucl. Instr. Meth. Phys. Res. B} {\bf 372} (2016)
  91 -- 101.

\bibitem{PerneggerFriedl2012}
H.~Pernegger and M.~Fiedl, {\it {Convoluted Landau and Gaussian Fitting
  Function}},  2012.
\newblock {based on Fortran code by R. Fruehwirth}.

\bibitem{Meroli2011}
S.~Meroli, D.~Passeri, and L.~Servoli, {\it Energy loss measurement for charged
  particles in very thin silicon layers},  {\em \jinst{6}{2011}{P06013}}
  (2011).

\bibitem{Laustriat1968}
G.~Laustriat, {\it The luminescence decay of organic scintillators},  {\em
  Molecular Crystals} {\bf 4} (1968) 127--145,
  [\href{http://arxiv.org/abs/http://dx.doi.org/10.1080/15421406808082905}{{\tt
  http://dx.doi.org/10.1080/15421406808082905}}].

\bibitem{Marradanetal2009}
T.~Marrod\'an~Undagoitia et~al., {\it Fluorescence decay-time constants in
  organic liquid scintillators},  {\em Review of Scientific Instruments} {\bf
  80} (2009) 043301.

\bibitem{Ranuccietal1994}
G.~Ranucci, P.~Ullucci, S.~Bonetti, I.~Manno, E.~Meroni, and A.~Preda, {\it
  Scintillation decay time and pulse shape discrimination of binary organic
  liquid scintillators for the {Borexino} detector},  {\em Nucl. Instr. Meth.
  Phys. Res. A} {\bf 350} (1994) 338 --350.

\bibitem{ErinPaper}
H.~O'Keeffe, E.~O'Sullivan, and M.~Chen, {\it Scintillation decay time and
  pulse shape discrimination in oxygenated and deoxygenated solutions of linear
  alkylbenzene for the {SNO+} experiment},  {\em Nucl. Instr. Meth. Phys. Res.
  A} {\bf 640} (2011) 119 -- 122.

\bibitem{Kumar1984}
C.~Kumar, S.~Chattopadhyay, and P.~Das, {\it Short-lived olefin triplets and
  energy transfer from them to $\beta$-carotene},  {\em Chemical Physics
  Letters} {\bf 106} (1984) 431 -- 436.

\bibitem{LaurelleMSc}
L.~Veloce, {\it {An Investigation of Backgrounds in the {DEAP-3600} Dark Matter
  Direct Detection Experiment}},  Master's thesis, Queen's University, 2013.

\bibitem{Amaudruz2015}
P.-A. Amaudruz et~al., {\it Radon backgrounds in the {DEAP-1}
  liquid-argon-based dark matter detector},  {\em Astroparticle Physics} {\bf
  62} (2015) 178 -- 194.

\bibitem{Francinietal2013}
R.~Francini et~al., {\it Tetraphenyl-butadiene films: {VUV-Vis} optical
  characterization from room to liquid argon temperature},  {\em
  \jinst{8}{2013}{C09010}} (2013).

\bibitem{Cavannapresentation}
F.~{Cavanna et al.}, {\it {VUV-VIS optical characterization of
  Tetraphenyl-butadiene files on glass and specular reflector substrates from
  room to liquid Argon temperature}},  2013.
\newblock {Light Detection in Noble Elements, Fermilab, Batvia, IL USA}.

\bibitem{Lyons1960}
L.~E. Lyons and J.~W. White, {\it 1003. luminescence from anthracene crystals
  and its temperature-dependence},  {\em J. Chem. Soc.} (1960) 5213--5218.

\bibitem{Schiffer1969}
H.-W. Schiffer, {\it {Temperaturabh\"angigkeit der Szintillationslichtausbeuten
  von p-Terphenyl-, Anthrazen- und Naphthalinkristallen bei Anregung mit
  $\alpha$- und $\beta$-Teilchen}},  {\em {Zeitschrift f\"ur Physik}} {\bf 227}
  (1969) 482--492.

\end{thebibliography}\endgroup

\end{document}